\documentclass[english,latin1,12pt]{article}
\usepackage[latin1]{inputenc}
\usepackage{float,color}
\usepackage{amsmath}
\usepackage{graphicx}
\usepackage{graphics,epsfig,natbib,}
\usepackage{setspace}
\usepackage{rotating}

\makeatletter 

\usepackage{amsmath,amssymb,amsfonts,latexsym,mathrsfs}
\usepackage{amsthm}
\usepackage{enumerate}
\usepackage{rotate}

\usepackage{latexsym}

\usepackage{babel}

\makeatother
\usepackage{fullpage}                   

\begin{document}
\newtheorem{theo}{Theorem}
\newtheorem{defi}{Definition}
\newtheorem{coro}{Corollary}
\newtheorem{prop}{Proposition}[section]
\newtheorem{rem}{Remark}
\newtheorem{lemma}{Lemma}
\newtheorem{proper}{Property}[section]
\newtheorem{algo}{Algorithm}

\newcommand{\tk}{P^{n} (x, dy)}
\newcommand{\bm}{\boldmath}
\newcommand{\bomega}{\boldsymbol{\omega}}
\newcommand{\bbeta}{\boldsymbol{\beta}}

\newcommand{\um}{\unboldmath}
\newcommand{\botheta}{\mbox{$\theta$}}
\newcommand{\boTheta}{\mbox{$\Theta$}}
\newcommand{\G}{\mbox{${\cal{G}}$}}
\newcommand{\M}{\mbox{${\cal{M}}$}}
\newcommand{\rcorr}{\mbox{corr}}
\newcommand{\rcov}{\mbox{cov}}
\newcommand{\RQMC}{\rm RQMC}
\newcommand{\rv}{\mbox{Var}}
\newcommand{\V}{\mbox{Var}}
\newcommand{\bolam}{\mbox{$\lambda$}}
\newcommand{\bopi}{\mbox{$p$}}
\newcommand{\ru}{{\mbox{Uniform}}}
\newcommand{\cas}{\mbox{${\cal{S}}$}}
\newcommand{\caf}{\mbox{${\cal{F}}$}}
\newcommand{\cag}{\mbox{${\cal{G}}$}}
\newcommand{\cac}{\mbox{${\cal{C}}$}}
\newcommand{\cau}{\mbox{${\cal{U}}$}}
\newcommand{\cae}{\mbox{${\cal{E}}$}}
\newcommand{\caz}{\mbox{${\cal{Z}}$}}
\newcommand{\cay}{\mbox{${\cal{Y}}$}}

\newcommand{\beq}{\begin{equation}}
\newcommand{\eeq}{\end{equation}}
\newcommand{\beqn}{\begin{eqnarray}}
\newcommand{\eeqn}{\end{eqnarray}}
\newcommand{\one}{{\mathbf 1}}
\newcommand{\RR}{{\mathbf R}}

\newcommand{\LL}{\mbox{${\cal{L}}$}}
\newcommand{\xe}{\mbox{E}}
\newcommand{\bu}{\mathbf{u}}
\newcommand{\bU}{\mathbf{U}}
\newcommand{\bW}{\mathbf{W}}
\newcommand{\bv}{\mathbf{v}}
\newcommand{\bw}{\mathbf{w}}
\newcommand{\bi}{\mathbf{i}}

\newcounter{example}[section]
\def\theexample{\thesection.\arabic{example}}
\setcounter{example}{0}

\renewenvironment{proof} {\noindent {\bf Proof } \it \bigskip}

\title{Embarrassingly Parallel Sequential Markov-chain Monte Carlo for Large Sets of Time Series
}

\author{Roberto Casarin\setcounter{footnote}{1}\footnotemark{}\hspace{15pt} %
Radu V. Craiu\setcounter{footnote}{2}\footnotemark{}\hspace{15pt} %
Fabrizio Leisen\setcounter{footnote}{3}\footnotemark{}\hspace{15pt} \\\\%
{\centering \small\setcounter{footnote}{1}\footnotemark{} University Ca' Foscari of Venice}\\%
{\centering \small\setcounter{footnote}{2}\footnotemark{} University of Toronto}\\
{\centering \small\setcounter{footnote}{3}\footnotemark{} University of Kent}%
}

\maketitle

\begin{abstract}
Bayesian computation crucially relies on Markov chain Monte Carlo (MCMC) algorithms. In the case of massive data sets, running  the Metropolis-Hastings sampler to draw from the posterior distribution becomes prohibitive due to the large number of likelihood terms that need to be calculated at each iteration. In order to perform Bayesian inference for a large set of  time series, we consider an algorithm that combines  ``divide and conquer'' ideas previously used to design MCMC algorithms for big data with a sequential MCMC strategy. The performance of the method is illustrated using a large set of financial data.    

\end{abstract}

{\it Key words:} Big Data, Panel of Time Series, Parallel Monte Carlo,  Sequential Markov-Chain Monte Carlo.

\section{Introduction}

There is little doubt that one of the main challenges brought on by  the advent of Big Data   in Bayesian statistics is to develop Markov chain Monte Carlo (MCMC) algorithms for  sampling a posterior distribution  derived from a very large sample. While MCMC has become the default tool to study  posterior distributions when they are not available in closed form,  many commonly used sampling algorithms, e.g. the  Metropolis-Hastings samplers, can become computationally prohibitive when  a large number of likelihood calculations are needed at each iteration.

In recent years we have witnessed a large research effort devoted to 
dividing the MCMC computational load among a number of available processors and recombining the results with as little loss in statistical efficiency as possible. For instance, the  approaches developed in  \cite{Scott2013} and \cite{NeiWanXin14}  divide the available data in smaller batches and sample
 the resulting {\it partial posteriors} obtained from each batch of data. They propose different methods  to combine the resulting draws  so that the efficiency of the resulting Monte Carlo estimator is close to the one that would have been obtained if the full data posterior had been sampled.  In the consensus MCMC of \cite{Scott2013} this is achieved via reweighting the partial posterior samples, while  the embarrassingly parallel approach of \cite{NeiWanXin14} relies on kernel density approximations of the partial posteriors to produce an approximation of the full one.  In \cite{WanDun13} the authors propose a refined recombination strategy based on the Weierstrass transformation of all partial posteriors. 

While dividing the whole data into batches can be done easily when the data are independent, one must proceed cautiously when the data exhibit long range dependencies, as is the case in time series. In such cases,  simply splitting time series into blocks  can lead to poor estimates of the parameters.  Instead, one can sometimes bypass the computational load by sequentially updating the posterior over time  \citep[see, for instance,][]{DunYan13}.  

Sequential  sampling may be improved when combined with parallel and possibly interacting  Monte Carlo methods that were used elsewhere, e.g. for   parallel adaptive MCMC \citep{craiu-jeff-yang},  for interacting MTM \citep{CasCraLei13}, for population Monte Carlo \citep{cappe04},  for Sequential Monte Carlo \citep{DelMoral:Miclo:00,DelMoral:04}  and for massively parallel computing \citep{LeeYauGilDouHol13} . 

Sequential estimation is  useful in many applied contexts such as on-line inference of econometric models for both out-of-sample and in-sample analyses. However, sequential estimation is a challenging issue in Bayesian analysis due to the computational cost of the numerical procedures for posterior approximation. Moreover, the computational cost rapidly increases with the dimension of the model and the inferential task becomes impossible. In this sense our paper contributes to the recent stream of the literature on the use of Central Processing Unit (CPU) and Graphics Processing Unit (GPU) parallel computing in econometrics (e.g., see \cite{DHS2002}, \cite{Swann2002}, \cite{C2005}, \cite{CG2008}, \cite{SHW2010}).

Our contribution here is to consider possible ways to combine  strategies following  the work of  \cite{NeiWanXin14}  with sequential MCMC in order to address difficulties that appear when studying  large panel time series data models.   Analyses of  panel time series data appear frequently in the  econometrics literature  as discussed in the review papers  of  \cite{canova} and  \cite{hsiao15}. {Moreover,  the use of latent variables for time series panel models in combination with a Bayesian inference approach \citep{kauf2,kauf,billio} can be quite challenging, due to the computational burden required for the latent variable estimation. These challenges motivate this work in which the  Bayesian stochastic volatility model recently proposed in \cite{WinCar14} and discussed in \cite{Cas14} is adapted to the context of large panel of time series.}

In  Sections 2 and 3 we introduce, respectively, the issues related to sequential sampling in models with latent variables and present our  algorithm. Section 4 contains the simulation studies and the real data analysis. The paper closes with conclusions and future directions.

\section{Posterior Distribution Factorization}\label{sec:model}

Consider a  time series sample $\mathbf{y}_{t}=(y_{t1},\ldots,y_{tm})\in Y \subset\mathbb{R}^{m}$, $t=1,\ldots,T$  with probability density function (pdf) $h_{t}(\mathbf{y}_{t}|\boldsymbol{\theta})$ where  $\boldsymbol{\theta}\in\Theta\subset\mathbb{R}^{p}$ is a parameter vector. Of interest is sampling from the posterior distribution of $\boldsymbol{\theta}$. We are considering here the case in which the latter task is made easier if  a data augmentation approach is adopted. We  introduce 
auxiliary variables $\mathbf{x}_{t}\in X\subset\mathbb{R}^{n}$, $1\le t \le T$ that exhibit Markovian serial dependence, i.e. each  $\mathbf{x}_{t}$ has pdf  $g(\mathbf{x}_{t}|\mathbf{x}_{t-1},\boldsymbol{\theta})$. If $f(\mathbf{y}_{t}|\mathbf{x}_{t},\boldsymbol{\theta})$ is the conditional pdf  of $\mathbf{y}_{t}$ given $\mathbf{x}_{t}$ and $\boldsymbol{\theta}$, then for prior distribution $\pi(\boldsymbol{\theta})$ we obtain the joint posterior of the data and the latent variables as
$$
\pi(\boldsymbol{\theta},\mathbf{x}_{1:T}|\mathbf{y}_{1:T})=\frac{1}{Z}\prod_{t=1}^{T}f(\mathbf{y}_{t}|\mathbf{x}_{t},\boldsymbol{\theta})g(\mathbf{x}_{t}| \mathbf{x}_{t-1},\boldsymbol{\theta})\pi(\boldsymbol{\theta})
$$
where $Z$ is the normalizing constant of $\pi(\boldsymbol{\theta},\mathbf{x}_{1:T}|\mathbf{y}_{1:T})$, $\mathbf{x}_{1:T}=\{\mathbf{x}_{1},\ldots,\mathbf{x}_{T}\}$ and   $\mathbf{y}_{1:T}=\{\mathbf{y}_{1},\ldots,\mathbf{y}_{T}\}$.

Henceforth, we  assume that the time series data has panel structure such that if we consider all the data collected up to time $t$, $\mathbf{y}_{1:t}$, then  it is possible to partition them into $M$ blocks of size $K$ each,
\beq
\mathbf{y}_{1:t} = \bigcup_{i=1}^{M} \mathbf{y}_{1:t}^{(i)}
\label{partition}
\eeq where the $i$th block $\mathbf{y}_{1:t}^{(i)}$ contains the measurements up to time $t$ for the components $k_{i-1},k_{i-1}+1,\ldots, k_{i}$ of $\mathbf{y}_{1:t}$ (for notational simplicity we set  $k_{i}=K \times i$ but other allocations are possible), i.e.  for all  $0\le i\le M$ 
$$\mathbf{y}_{1:t}^{(i)}=\mathop{\bigcup} \limits_{j=k_{i-1}+1}^{k_{i}} \{ y_{1j} ,\ldots, y_{tj}\} := \mathop{\bigcup} \limits_{j=k_{i-1}+1}^{k_{i}}  \mathbf{y}_{1:t,j}\;.$$  Each set $\mathbf{y}_{1:t}^{(i)}$  of the partition contains  the $i$th panel of dependent components  of   $\mathbf{y}_{t}$. 
An important assumption of the model is that, conditional on a parameter value $\boldsymbol{\theta}$, the components in two partition sets are independent, i.e. \beq
\mathbf{y}_{1:t}^{(i)} \perp\mathbf{y}_{1:t}^{(i')},
 \label{A1}
 \eeq
  for any $i \ne i'$.  

Corresponding to the partition \eqref{partition} there is an equivalent partition of the auxiliary variables 
\beq
\mathbf{x}_{1:t}=\mathop{\bigcup} \limits_{i=1}^{M} \mathbf{x}_{1:t}^{(i)},
\label{partition2}
\eeq
where the components of $\mathbf{x}_{1:t}^{(i)}$ correspond to the components of $\mathbf{y}_{t}$ included in $\mathbf{y}_{1:t}^{(i)}$. A second crucial assumption for the validity of our algorithm is the independence of the  auxiliary variables contained in two elements of the partition \eqref{partition2}, i.e.  \beq \mathbf{x}_{1:t}^{(i)} \perp \mathbf{x}_{1:t}^{(i')}, \label{A2}
\eeq
for all $1\le i \ne i' \le M$.  Finally, we assume that $\mathbf{y}_{1:t}^{(i)}$ depends  only on those auxiliary variables included in 
$\mathbf{x}_{1:t}^{(i)}$, i.e. given  the latter we have 
\beq
\mathbf{y}_{1:t}^{(i)}\perp \mathbf{x}_{1:t}^{(i')}.
\label{partition3}
\eeq

These assumptions are not unusual in the context of dynamic panel data models  as they are used for theoretical derivations in \cite{atw}, \cite{blund} and \cite{bone} as well as in applications \citep[see, for instance,][]{abrev,kauf2,kauf,billio}.


The basic principle underlying our approach is to  learn sequentially over time using different cross-sectional data blocks.  Our algorithm relies on samples from the the joint posterior distribution of $\boldsymbol{\theta}$ and $\mathbf{x}_{1:t}^{(i)}$, $\pi_{it}(\boldsymbol{\theta},\mathbf{x}_{1:t}^{(i)})$, conditional on the sub-sample $\mathbf{y}_{1:t}^{(i)}$ 
\begin{equation}
\pi_{it}(\boldsymbol{\theta},\mathbf{x}_{1:t}^{(i)})=\pi(\boldsymbol{\theta},\mathbf{x}_{1:t}^{(i)}|\mathbf{y}_{1:t}^{(i)})
=\frac{1}{Z_{it}}\prod_{s=1}^{t}f(\mathbf{y}_{s}^{(i)}|\mathbf{x}_{s}^{(i)},\boldsymbol{\theta})g(\mathbf{x}_{s}^{(i)}|\mathbf{x}_{s-1}^{(i)},\boldsymbol{\theta})\pi(\boldsymbol{\theta})^{1/M}
\end{equation}
where $\mathbf{y}_{s}^{(i)} = \{ y_{s{k_{i-1}}+1},\ldots,y_{s{k_{i}}} \}$,  
$\mathbf{x}_{s}^{(i)} = \{ x_{sk_{i-1}+1},\ldots,x_{s{k_{i}}} \}$ and $Z_{it}$ is the $i$-th block normalizing constant. 

Using the assumptions  \eqref{A1} and \eqref{A2} it results that 
\begin{eqnarray}
\pi(\boldsymbol{\theta},\mathbf{x}_{1:t}|\mathbf{y}_{1:t})&\propto&   \pi(\boldsymbol{\theta}) f(\mathbf{y}_{1:t}|\mathbf{x}_{1:t},\boldsymbol{\theta}) g(\mathbf{x}_{1:t}|\boldsymbol{\theta})= \nonumber \\
&=& [\pi(\boldsymbol{\theta})^{1/M}]^{M} \prod_{i=1}^{M} f(\mathbf{y}_{1:t}^{(i)}|\mathbf{x}_{1:t}^{(i)},\boldsymbol{\theta}) g(\mathbf{x}_{1:t}^{(i)}|\boldsymbol{\theta}) \propto \prod_{i=1}^{M}\pi_{it}(\boldsymbol{\theta},\mathbf{x}_{1:t}^{(i)}|\mathbf{y}_{1:t}^{(i)})
 \label{joint-factorization}
 \end{eqnarray}

From \eqref{joint-factorization} we can infer that the type of factorization of the posterior distribution used in \cite{NeiWanXin14} holds in this case for every $t$ since
 
\begin{eqnarray}
\pi(\boldsymbol{\theta} |\mathbf{y}_{1:t}) &\propto& \int \pi(\boldsymbol{\theta} , \mathbf{x}_{1:t} |\mathbf{y}_{1:t})  d\mathbf{x}_{1:t}  \nonumber\\
&\propto & \int \ldots \int \prod_{i=1}^{M} \pi_{it}(\boldsymbol{\theta},\mathbf{x}_{1:t}^{(i)}|\mathbf{y}_{1:t}^{(i)})  d\mathbf{x}_{1:t}^{(1)} \ldots d\mathbf{x}_{1:t}^{(M)} = \prod_{i=1}^{M}
\pi_{it}(\boldsymbol{\theta} |\mathbf{y}_{1:t}^{(i)}). \label{joint-posterior}
 \end{eqnarray}


\section{Embarrassingly Parallel SMCMC}\label{sec:algorithm}
So far we have discussed the factorization of the posterior distribution based on the panel-type  structure of the data. In this section we show how the algorithm  handles the serial dependence in the data and samples from $\{\pi_{it}:\; 1\le t \le T\}$ for each $i\in \{1,\ldots,M\}$ using  a sequential MCMC strategy. For expository purposes we assume that the  parameter estimates are updated every time a new observation become available, but in the application we will update the estimates after every  $J$th observation is collected.

Let us define $\boldsymbol{\lambda}_{t}=(\boldsymbol{\theta},\mathbf{x}_{1:t})$, $t\in\mathbb{N}$, the time sequence of augmented parameter vectors with non-decreasing dimension $d_{t}=d_{t-1}+d$, $t\geq 1$. In order to take advantage of  the partition described in the previous section we also introduce $\boldsymbol{\lambda}_{t}^{(i)}=(\boldsymbol{\theta},\mathbf{x}_{1:t}^{(i)})$,  a parameter vector of dimension $d_{t}^{(i)}$. Since the augmented parameter vector can then be partitioned as $\boldsymbol{\lambda}_{t}=(\boldsymbol{\lambda}_{t-1},\mathbf{x}_{t})$ and  $\boldsymbol{\lambda}_{t}^{(i)}=(\boldsymbol{\lambda}_{t-1}^{(i)},\mathbf{x}_{t}^{(i)})$ for all $1\le i \le M$, the correctness of the algorithms relies on the compatibility condition for the priors on $\boldsymbol{\lambda}_{t}^{(i)}$ at each time $t$ and for each $i$th block of data. Specifically,  we assume that if the prior is 
$p_{t}^{(i)}(\boldsymbol{\lambda}_{t}^{(i)})=p(\boldsymbol{\theta})p(\mathbf{x}_{1:t}^{(i)}|\boldsymbol{\theta})$, where 
$p(\mathbf{x}_{1:t}^{(i)}|\boldsymbol{\theta})=\prod_{t=1}^{T}g(\mathbf{x}_{t}^{(i)}|\mathbf{x}_{t-1}^{(i)},\boldsymbol{\theta})$, then it satisfies the compatibility condition
\begin{equation}
p_{t}^{(i)}(\boldsymbol{\lambda}_{t}^{(i)})=\int p_{t+1}^{(i)}(\boldsymbol{\lambda}_{t}^{(i)},\mathbf{x}_{t+1}^{(i)})d\mathbf{x}_{t+1}^{(i)}, \; \forall 1\le i \le M, \; 1\le t \le T-1.  
\label{A3}
\end{equation}

Our embarrassing SMCMC algorithm iterates over time and data blocks. At each time $t=1,\ldots,T$, the algorithm consists of two steps. In the first step, for each data block $\mathbf{y}_{1:t}^{(i)}$, $i=1,\ldots,M$ we use  $L$ parallel SMCMC chains, each of which yields $n_{t}$ samples from $\pi_{it}(\boldsymbol{\lambda})$, i.e. we generate 
$\boldsymbol{\lambda}_{it}^{(l,j)}$, $j=1,\ldots,n_{t}$, and $l=1,\ldots,L$ from $\pi_{it}(\boldsymbol{\lambda})$. Based on {\it all} samples $\boldsymbol{\theta}_{it}^{(l,j)}$, $j=1,\ldots,n_{t}$, and $l=1,\ldots,L$ we produce the kernel density estimates 
$\hat{\pi}_{it}(\boldsymbol{\theta})$ of the marginal  sub-posteriors $\pi_{it}(\boldsymbol{\theta})$. 

In the second step we take advantage of the factorization  \eqref{joint-factorization} and the 
 posterior $\pi(\boldsymbol{\theta}| \mathbf{y}_{1:T})$ is approximated by combining the approximate sub-posteriors $\hat{\pi}_{iT}(\boldsymbol{\theta})$, $i=1,\ldots,M$. Samples from this distribution can be obtained by applying the asymptotically exact posterior sampling procedure detailed in Algorithm 1 of \cite{NeiWanXin14}. It is worth noting that \eqref{joint-factorization}  holds for any $t=1,\ldots,T$ so, if needed, one can approximate $\pi(\boldsymbol{\theta}| \mathbf{y}_{1:t})$ at intermediate times $t \in \{1,\ldots, T\}$ by combining the $\hat{\pi}_{it}$'s. However, $\pi(\boldsymbol{\theta}| \mathbf{y}_{1:t})$  is not used directly in the final posterior    $\pi(\boldsymbol{\theta}| \mathbf{y}_{1:T})$ so if one is interested only in the latter then the merging of partial posteriors is performed only at time $T$.

The pseudocode of the proposed EP-SMCMC is given in Algorithm \ref{alg1} and the details of the SMCMC and of the merge step are detailed in the following sections.
   

\begin{center}
\begin{minipage}[thp]{340pt} \par\hrule\vspace{5pt}
\begin{algo}{Embarrassingly Parallel SMCMC (EP-SMCMC)}\label{alg1}
\par\vspace{5pt}\hrule
\par\vspace{5pt} For $t=1,2,\ldots,T$
\begin{enumerate}
\item For $i=1,\ldots,M$ 
draw $\boldsymbol{\lambda}_{it}^{(l,j)}$, $j=1,\ldots,n_{t}$, and $l=1,\ldots,L$ from $\pi_{it}(\boldsymbol{\lambda})$ by using the SMCMC transition.

\item When needed (usually at time $t=T$):
\begin{enumerate}
\item[(a)] Compute the kernel density estimate $\hat{\pi}_{it}(\boldsymbol{\theta})$ of the marginal sub-posteriors $\pi_{it}(\boldsymbol{\theta})$ by using the samples $\boldsymbol{\theta}_{it}^{(l,j)}$, $j=1,\ldots,n_{t}$, and $l=1,\ldots,L$.

\item[(b)] approximate the posterior $\pi(\theta| \mathbf{y}_{1:t})$ by combining the approximate sub-posteriors $\hat{\pi}_{it}(\boldsymbol{\theta})$, $i=1,\ldots,M$.
\end{enumerate}

\end{enumerate}
\end{algo}
\hrule\vspace{5pt}
\end{minipage}
\end{center}

\subsection{Sequential MCMC}
In this section we discuss the construction of the SMCMC samplers that are used in the first step of the EP-SMCMC. To simplify the notation we drop the index $i$ indicative of the data block. 
In the SMCMC algorithm a population of $L$ parallel inhomogeneous Markov chains are used to generate the samples $\boldsymbol{\lambda}_{t}^{(l,j)}$ with $j=1,\ldots,n_{t}$, $l=1,\ldots,L$ and $t=1,\ldots,T$ from the sequence of posterior distributions $\pi_{t}$, $t=1,\ldots,T$. Each Markov chain of the population is defined by a sequence of transition kernels $K_{t}(\boldsymbol{\lambda},A)$, $t\in\mathbb{N}$, that are  operators from $(\mathbb{R}^{d}_{t-1},\mathcal{B}(\mathbb{R}^{d}_{t-1}))$ to $(\mathbb{R}^{d}_{t},\mathcal{B}(\mathbb{R}^{d}_{t}))$, such that $K_{t}(\boldsymbol{\lambda},\cdot)$ is a probability measure for all $\boldsymbol{\lambda}\in\mathbb{R}^{d_{t-1}}$, and $K_{t}(\cdot,A)$ is measurable for all $A\in\mathcal{B}(\mathbb{R}^{d_{t}})$. 

The kernel $K_{t}(\boldsymbol{\lambda},A)$ has $\pi_{t}$ as stationary distribution and results from the composition of a jumping kernel, $J_{t}$ and a transition kernel, $T_{t}$, that is
$$
K_{t}(\boldsymbol{\lambda},A)=J_{t}\circ T_{t}^{n_{t}}(\boldsymbol{\lambda},A)=\int_{\mathbb{R}^{d_{t}}}J_{t}(\boldsymbol{\lambda},d\boldsymbol{\lambda}')T_{t}^{n_{t}}(\boldsymbol{\lambda}',A)
$$
where the fixed dimension transition is defined as 
$$
T_{t}^{m_{t}}(\boldsymbol{\lambda},A)=T_{t}\circ T_{t}^{n_{t}-1}(\boldsymbol{\lambda},A)=\int_{\mathbb{R}^{d_{t}}}T_{t}(\boldsymbol{\lambda},d\boldsymbol{\lambda}')T_{t}^{n_{t}-1}(\boldsymbol{\lambda}',A)
$$
with $n_{t}\in\mathbb{N}$, and $T^{0}=Id$ is the identity kernel. We assume that the jumping kernel satisfies 
$$
J_{t+1}(\boldsymbol{\lambda}_{t},\boldsymbol{\lambda}_{t+1})=J_{t+1}(\boldsymbol{\lambda}_{t},\mathbf{x}_{t+1})\delta_{\boldsymbol{\lambda}_{t}}(\tilde{\boldsymbol{\lambda}}_{t}),
$$
where $\boldsymbol{\lambda}_{t+1}=(\tilde{\boldsymbol{\lambda}}_{t},\mathbf{x}_{t+1})$ and $J_{t+1}(\boldsymbol{\lambda}_{t},\mathbf{x}_{t+1})=J_{t+1}(\boldsymbol{\lambda}_{t},(\boldsymbol{\lambda}_{t},\mathbf{x}_{t+1}))$.
This condition ensures that the error propagation through the jumping kernel can be controlled over the SMCMC iterations.  

In order to apply the SMCMC one need to specify the transition kernel $T_{t+1}$ and the jumping kernel $J_{t+1}$ at the iteration $t+1$. The transition kernel $T_{t}$ at the iteration $t$ allows each parallel chain to explore the sample space of a given dimension, $d_t$, and to generate samples $\boldsymbol{\lambda}_{t}^{(l,j)}$, from the posterior distribution  $\pi_t$. The jumping kernel $J_{t+1}$ at the iteration $t+1$, allows the chains to go from a space of dimension $d_{t}$ to one of dimension $d_{t+1}$. 

\subsection{Merge step}
The merge step relies on the following approximation the posterior distribution 
\begin{equation}
\pi_{t}(\boldsymbol{\theta})=\prod_{i=1}^{M}\hat{\pi}_{it}(\boldsymbol{\theta})
\end{equation}
where
\begin{eqnarray}
\hat{\pi}_{it}(\boldsymbol{\theta})&=&\frac{1}{N_{t}}\sum_{l=1}^{L}\sum_{j=1}^{n_{t}}\frac{1}{h^{p}}K\left(\frac{||\boldsymbol{\theta}-\boldsymbol{\theta}_{it}^{(l,j)}||}{h}\right)\nonumber\\
&=&\frac{1}{N_{t}}\sum_{j_{i}=1}^{N_{t}}\frac{1}{h^{p}}K\left(\frac{||\boldsymbol{\theta}-\boldsymbol{\theta}_{it}^{(j_{i})}||}{h}\right)
\end{eqnarray}
where $\boldsymbol{\theta}_{it}^{(k)}=\boldsymbol{\theta}_{it}^{(l,j)}$ with $l=k\, \hbox{div}\, L +1$ and $j=k\,\hbox{mod}\, L$, $N_{t}=n_{t}L$, and $K$ is a Gaussian kernel with bandwidth parameter $h$. Following the embarrassing MCMC approach the posterior distribution can be written as
\begin{equation}
\pi_{t}(\boldsymbol{\theta})\propto\sum_{j_{1}=1}^{N_{t}}\ldots\sum_{j_{m}=1}^{N_{t}}w_{j\cdot}\frac{1}{\bar{h}^p}K\left(\frac{\boldsymbol{\theta}-\bar{\boldsymbol{\theta}}_{j\cdot}}{\bar{h}}\right),
\end{equation}
where $\bar{h}=h/\sqrt{M}$, and 
$$
\bar{\boldsymbol{\theta}}_{j\cdot}=\frac{1}{M}\sum_{i=1}^{M}\boldsymbol{\theta}_{it}^{(j_{i})},\quad\quad
w_{j\cdot}=\prod_{i=1}^{M}\frac{1}{\bar{h}^{p}}K\left(\frac{||\boldsymbol{\theta}_{it}^{(j_i)}-\bar{\boldsymbol{\theta}}_{t\cdot}||}{h}\right).
$$

\subsection{Parameter tuning}
\noindent As suggested by \cite{DunYan13}, the number of iterations of the Sequential MCMC sampler at each time $n_t$ are chosen accordingly to  the correlation across the parallel chains. We let the number of iterations at the iteration $t$, $n_t$, be the smallest integer $s$ such that $r_t\left(s\right)\leq1-\epsilon$, where $r_t\left(s\right)$ is the rate function associated with the transition kernel $T_t$ and $\epsilon$ is a given threshold level. An upper bound for the rate function is provided by chain autocorrelation function at the $s$-th lag. It can be estimated sequentially using the output of all the parallel chains: $\hat{r}_t\left(s\right)=\max\{\hat{r}(s,j),j=1,\ldots,(n_t+p) \}$ where
\begin{eqnarray}
\hat{r}(s,j)=\frac{\sum_{l=1}^L\left(\lambda_j^{(s+1,t,l)}-\bar{\lambda}_j^{(s+1,t)}\right)
\left(\lambda_j^{(1,t,l)}-\bar{\lambda}_j^{(1,t)}\right)}{\left(\sum_{l=1}^L\left(\lambda_j^{(s+1,t,l)}-\bar{\lambda}_j^{(s+1,t)}\right)^2\right)^{\frac{1}{2}}
\left(\sum_{l=1}^L\left(\lambda_j^{(1,t,l)}-\bar{\lambda}_j^{(1,t)}\right)^2\right)^{\frac{1}{2}}},
\end{eqnarray}
with $\lambda_j^{(s,t,l)}$ the $j$-th element of the vector $\boldsymbol{\lambda}^{(s,t,l)}$ of the parameters $\boldsymbol{\gamma}^{(l)}$ and the latent states generated up to time $t$ by the $l$--th chain at the the $s$--th iteration. $\bar{\lambda}_j^{(s,t)}=L^{-1}\sum_{l=1}^L\lambda_j^{(s,t,l)}$ is the average of the draws over the $L$ parallel chains.

\section{Numerical Experiments}\label{sec:results}
\subsection{Simulated data}
{We consider a time series model in which  $y_{it}$ is the realized variance for the $i$-th log-return series at time $t$, and $x_{it}$ is the latent stochastic volatility process. In order to capture the time variations in the volatility of the series we consider the exponential family state space model for positive observations recently proposed in \cite{WinCar14} and extend it to the context of panel data.} The model can be mathematically described then as:
\begin{eqnarray}
y_{it}|x_{it}&\sim&\mathcal{G}a\left(\kappa/2,\kappa x_{it}/2\right)\label{obsbis}\\
x_{it}|x_{it-1}&\sim& x_{it-1}\psi_{it}/\lambda,\quad\psi_{it}\sim\mathcal{B}e(\nu/2,\kappa/2)\label{obsbis1}
\end{eqnarray}
for $t=1,\ldots,T$, where $\mathcal{G}a(a,b)$ denotes the gamma distribution and $\mathcal{B}e(a,b)$ the beta distribution of the first type. 

\begin{figure}[t]
\begin{center}
\begin{tabular}{c}
\vspace{-60pt}\\
\includegraphics[width=0.9\textwidth, height=400pt]{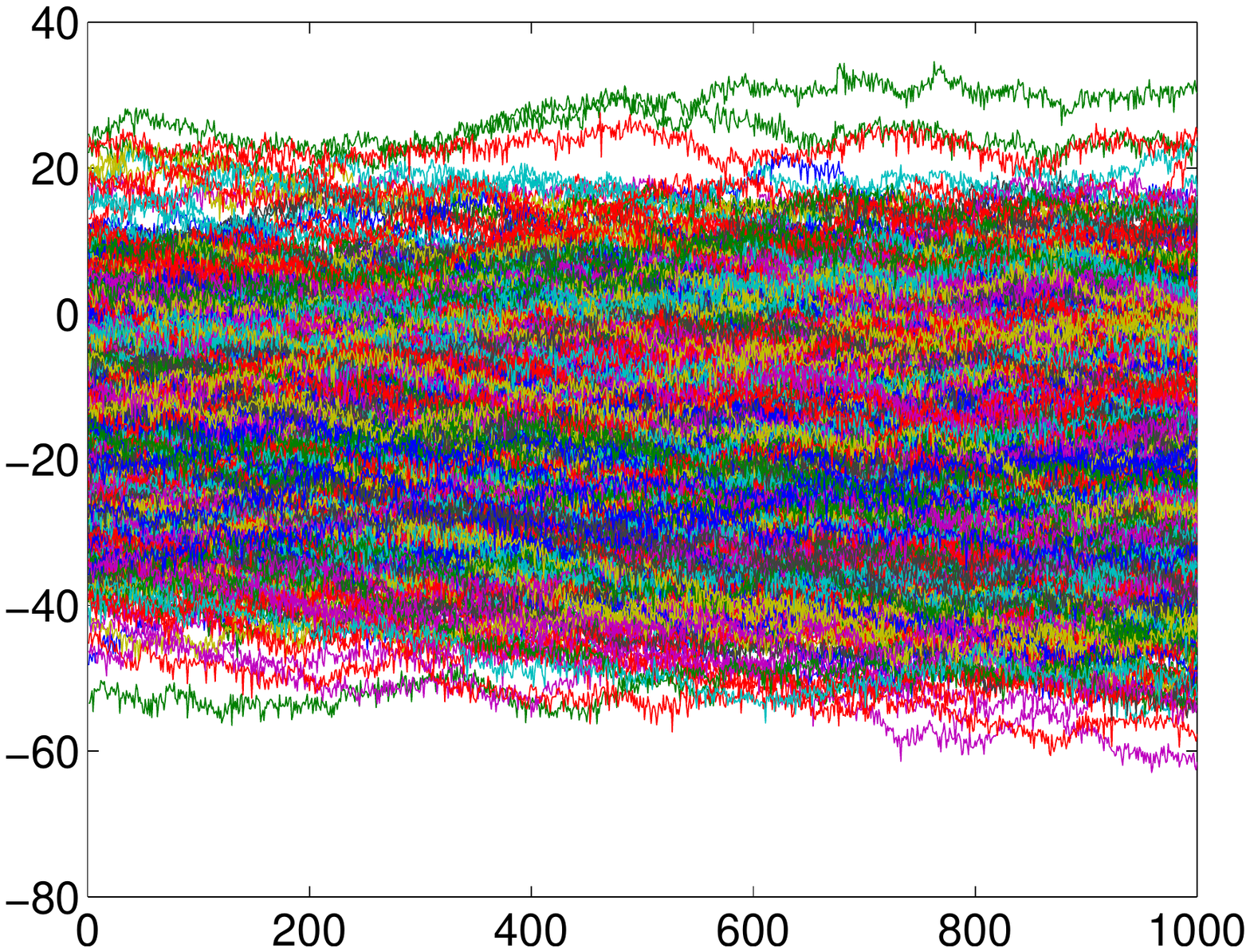}
\end{tabular}
\end{center}
\vspace{-90pt}
\caption{Samples $y_{it}$ (in log-scale), $i=1,\ldots,m$, $t=1,\ldots,T$, simulated from the state-space model in Eq. \ref{obsbis}-\ref{obsbis1}, with $T=1,000$, $m=1,000$ and parameter setting $\lambda=0.7$, $\kappa=3.8$ and $\nu=10$.}\label{SimData}
\end{figure}

We generate 1000 time series of 1000 observations each, and obtain a dataset of 1 million observations. In Fig. \ref{SimData} we illustrate such a simulated dataset. Inference for a nonlinear latent variable model with this large a sample  via MCMC sampling can be computationally challenging. In the simulation we set $\lambda=0.7$, $\kappa=3.8$ and $\nu=10$, with initial condition $x_{0i}=10$, $\forall i$. For each series  we have  generated 3,000 realizations, but the initial 2000 samples were discarded so that dependence on initial conditions can be considered negligible.

We aim to estimate the common parameters $\lambda$, $n$ and $\kappa$ and assume a uniform prior distribution for $\lambda$, i.e. $\lambda\sim\mathcal{U}{[0,1]}$ and  proper vague prior distributions for $\nu$ and $\kappa$, that is $\nu\sim\mathcal{G}a(0.5,0.5)$ and $\kappa\sim\mathcal{G}a(0.5,0.5)$ truncated on the region $\{(\nu,\kappa): \; \nu>\kappa-1\}$.

In order to apply the EP-SMCMC algorithm we assume that the $m$ series are split into $M$ blocks of $K=m/M$ series each.  The updates are performed  every $J$ observations, so the total number of updates is $n=T/J$. Let us denote with the column vector $\mathbf{u}_{1:t}=(u_{s},\ldots,u_{t})'$ a collection of variables $u_{r}$ with $r=s,\ldots,t$.  We define the $i$-th block of series and the $i$-th block of latent variables as the $(t\times K)$-matrices $Y_{it}=(\mathbf{y}_{(i-1)K+1,1:t},\ldots,\mathbf{y}_{iK,1:t})$ and $X_{it}=(\mathbf{x}_{(i-1)K+1,1:t},\ldots,\mathbf{x}_{iK,1:t})$, respectively, with $\mathbf{y}_{j,1:t}=(y_{j1},\ldots,y_{jt})'$, and $\mathbf{x}_{j,1:t}=(x_{j1},\ldots,x_{jt})'$. Then, the complete-data likelihood function at time $t$ for the $i$-th block is
\begin{align}
&L(Y_{it},X_{it}|\boldsymbol{\theta})=\prod_{j=k_{i-1}+1}^{k_i}\prod_{s=1}^{t}\frac{1}{\Gamma(\frac{\kappa}{2})}\left(\frac{\kappa x_{js}}{2}\right)^{\kappa/2} y_{js}^{\frac{\kappa}{2}-1}\exp\left(-\frac{\kappa x_{js}}{2} y_{js}\right)\\
&\quad \left(\frac{\lambda x_{js}}{x_{js-1}}\right)^{\frac{n}{2}-1}\left(1-\frac{\lambda x_{js}}{x_{js-1}}\right)^{\frac{\kappa}{2}-1}\frac{\lambda}{x_{js-1}}
\end{align}
where $\boldsymbol{\theta}=(\lambda,\kappa,\nu)$. Then the sub-sample posterior, based on the complete-data likelihood function of the $i$-th block is 
\begin{equation}
\pi(\boldsymbol{\theta},X_{it}|Y_{it})\propto L(Y_{it},X_{it}|\boldsymbol{\theta})(\nu+\kappa)^{-\frac{M}{2}}\exp\left(-\frac{M}{2}(\nu+\kappa)\right)\mathbb{I}(\nu>\kappa-1)
\end{equation}

At time $t$, and for the $i$-th block, the $l$-th SMCMC parallel chain has the transition kernel of a Gibbs sampler which iterates over the following steps:
\begin{itemize}
\item[1.1] generate $\boldsymbol{\theta}^{(l,j)}$ from $f(\boldsymbol{\theta}|Y_{it},X_{it}^{(l,j-1)})$
\item[1.2] generate $X_{it}^{(l,j)}$ from $f(X_{it}|\boldsymbol{\theta}^{(l,j)},Y_{it})$
\end{itemize}
with $j=1,\ldots,n_t$, and $X_{it}^{(l,0)}=((X_{it-1}^{(l,n_{t-1})})',(x_{k_{i-1}+1,t}^{(l,1)},\ldots,x_{k_{i},t}^{(l,1)})')'$ is a $(t\times K) $-dim matrix where the $t$-th row elements drawn from the jumping kernel at time $t-1$. At time $t+1$, as a new observation become available, the dimension of the state space for the $i$-th block SMCMC chains increase from $d_t$ to $d_{t+1}=d_t+J$. We choose as jumping kernel of the $l$-th parallel chain to be the transition kernel of a Gibbs sampler with the following full conditional distribution
\begin{itemize}
\item[2.] $x_{kt+1}^{(l,1)}\sim f(x_{kt+1}|X_{it}^{(l,j)},X_{it}^{(l,j)},\boldsymbol{\theta}^{(l,j)})$, with $k=k_{i-1}+1,\ldots,k_{i}$
\end{itemize}
where $j=n_{t}$. The details of the sampling procedures for the three steps are given in Appendix \ref{app}.

In the simulation example we compare our EP-SMCMC with a MCMC repeated sequentially over time and a Sequential Monte Carlo (SMC) with unknown parameters. For the EP-SMCMC we used $L=10$ parallel chains and a number of iterations $n_t$ close to 50, on average. For the MCMC we use a multi-move Gibbs sampler where the latent states are sampled in one block by applying a forward-filtering backward sampling (FFBS) procedure. The filtering and smoothing relationships are given in Appendix \ref{appSMCMC}. In order to update the parameters we have used a Metropolis-Hastings step. The MCMC chain of our multi-move sampler is mixing quite well due to the FFBS and also the Metropolis step has acceptance rates about 0.3 which is a good rate for many models as argued in \cite{MR1888450}.

In our MCMC analysis we considered two cases. The first one is based on samples of 1,000 iterations after a burn-in phase of 500 iterations in order to have a computational complexity similar to the one of the EP-SMCMC. The second one is based on samples of 25,000 iterations after a burn-in phase of 20,000 iterations and is used to have reliable MCMC estimates of the parameter posterior distribution based on the whole sample.  

For the SMC we also consider the regularized auxiliary particle filter (RAFP) combined with the embarrassingly parallel algorithm to obtain a EP-RAPF. In Appendix \ref{AppSMC} we present  the computational details of running the r-APF   for our stochastic volatility model.  We refer to  \cite{LiuWes01} for the definition of regularized particle filter and to \cite{Casarin:Marin:09} for a comparison of different regularized filters for stochastic volatility models.

When we  compare the EP-SMCMC, MCMC and EP-RAPF algorithms,  we use as a measure of efficiency the mean square errors for the parameters $\kappa$, $\nu$ and $\lambda$ after the last time-block of data has been processed. The mean square errors have been estimated using 40 independent runs of each algorithm. The same set of data has been used to reduce the standard error of the MSE estimates. The experiments have been conducted on a cluster multiprocessor system with 4 nodes; each node comprises of four Xeon E5-4610 v2 2.3GHz CPUs, with 8 cores, 256GB ECC PC3-12800R RAM, Ethernet 10Gbit, 20TB hard disk system with Linux. The algorithms have been implemented in Matlab (see \cite{MATLAB}) and the parallel computing makes use of the Matlab parallel computing toolbox.

\begin{figure}[p]
\begin{center}
\vspace{-60pt}
\begin{tabular}{c}
\includegraphics[width=0.85\textwidth, height=280pt]{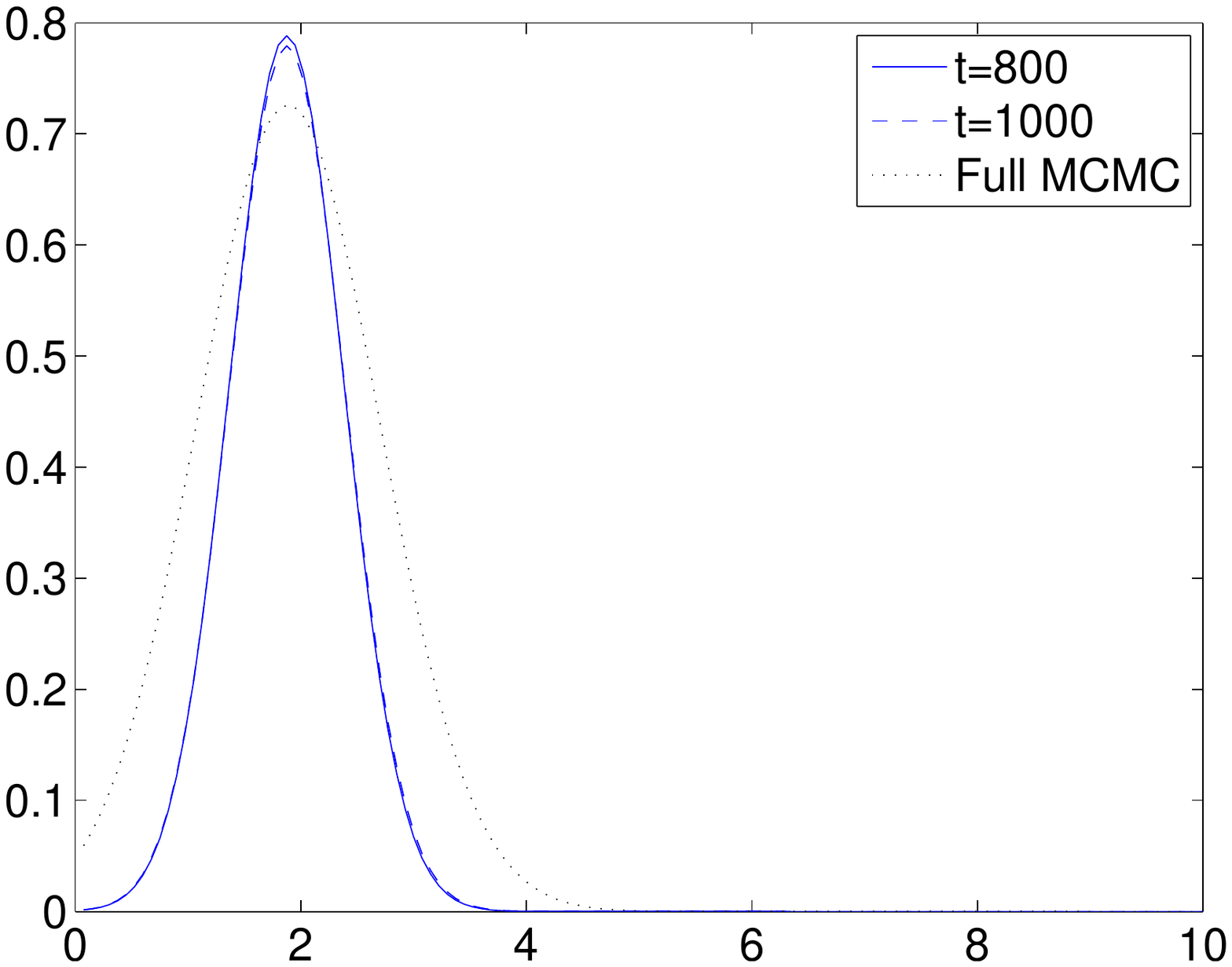}\\
\vspace{-150pt}\\
\includegraphics[width=0.85\textwidth, height=280pt]{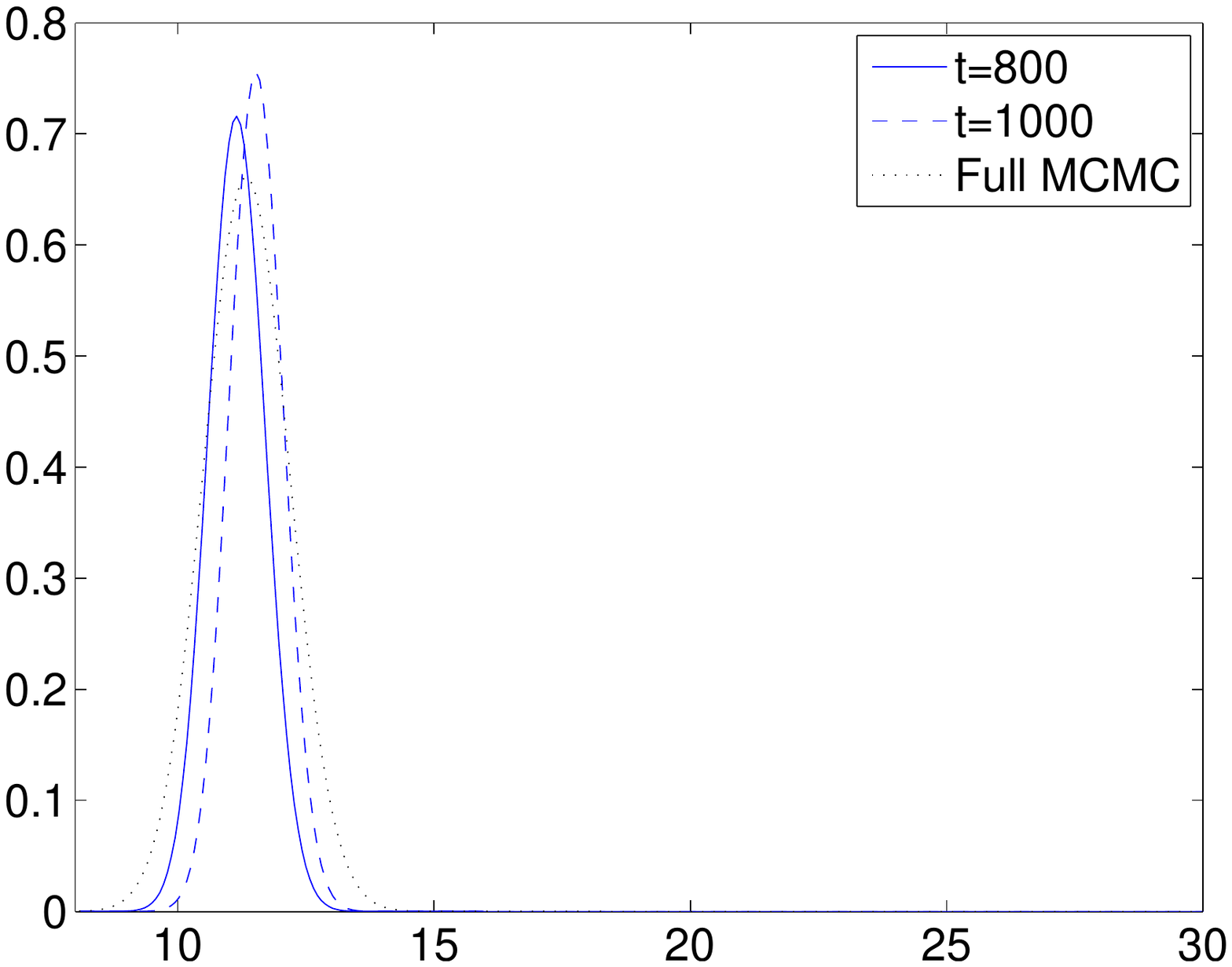}\\
\vspace{-150pt}\\
\includegraphics[width=0.85\textwidth, height=280pt]{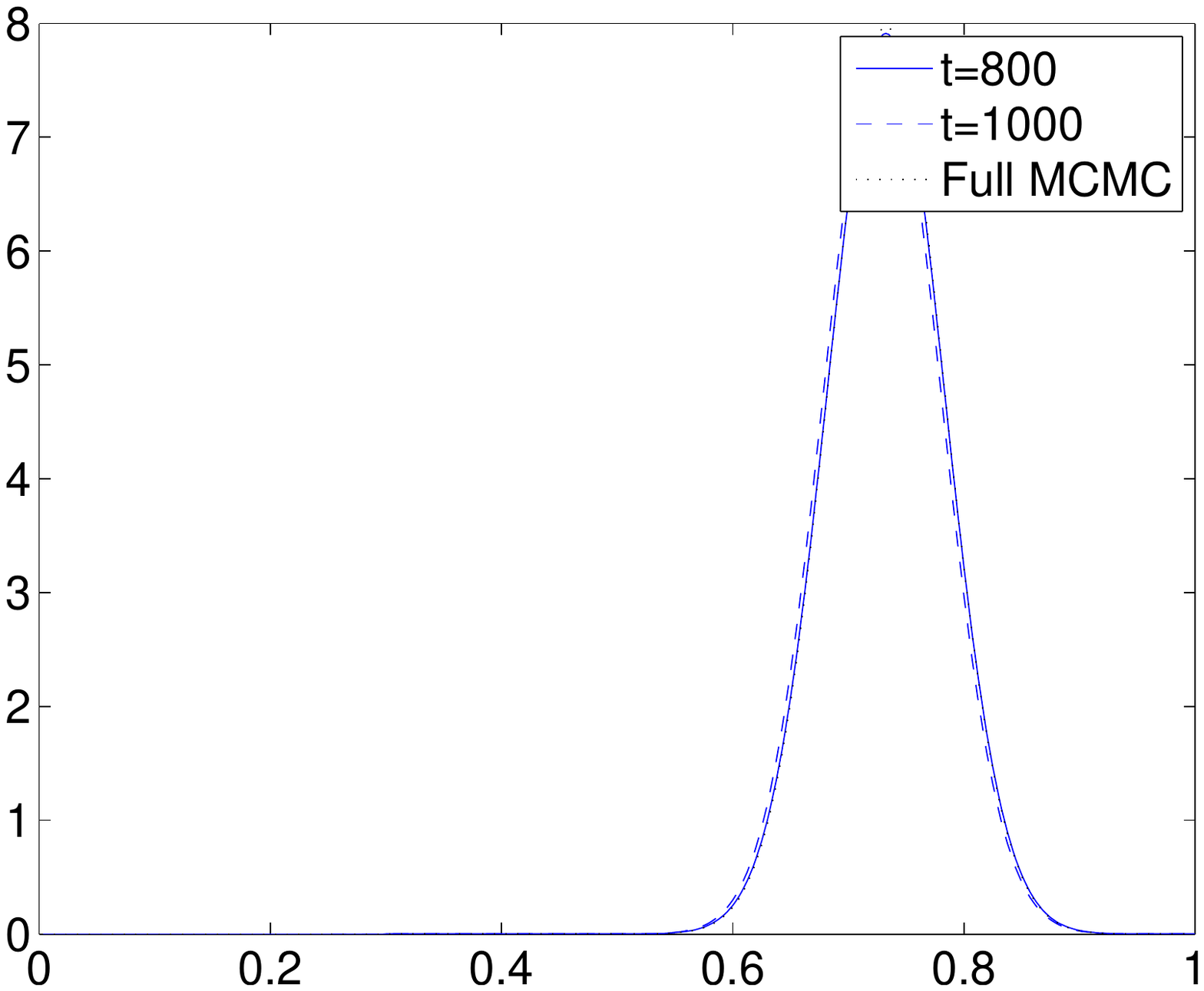}
\end{tabular}
\end{center}
\vspace{-50pt}
\caption{Estimation of the posterior densities of the parameters $\kappa$ (top panel), $\nu$ (middle panel) and $\lambda$ (bottom panel).  The EP-SMCMC  estimates are obtained at   iterations 800 (solid line) and 1000 (dashed line) when  $K=30$ and $J=200$ . The full posterior densities estimates are obtained from a MCMC with 25,000 iterations  (dotted line).}\label{SimPost}
\end{figure}

{The structure of the EP-SMCMC sampler allows for sequential data acquisition and for parallel estimation based on different cross-sectional blocks. Thus, the MSE obtained after processing the last block is given in Table \ref{TabMSE} for different dimensions of the time blocks $J$ (rows) and of the cross section blocks $M$ (columns). Fig. \ref{SimPost} shows the posterior approximation obtained from one run of the EP-SMCM on the dataset shown in Fig. \ref{SimData}. From our experiments we find that the parameters that are most difficult to estimate are $\kappa$ and $\nu$, whereas $\lambda$ has lower MSEs regardless of the choice of  block size and type of algorithm. For the EP-SMCMC sampler a larger  size $J$ of the time blocks reduces the MSE,  possibly due to a reduction in the propagation of the approximation error over the iterations. The behaviour of the MSE with respect to the cross-sectional block size $K$ is not monotonic. The MSE initially decreases with $K$, but for larger $K$ it increases. From our experiments, values of $K$ between 20 and 40 yield the best results. The $\kappa$ and $\nu$ MSEs for the MCMC are higher then their EP-SMCMC counterparts, likely due to the lack of convergence of the MCMC in the 1,500 iterations per time block. The large MSEs for the EP-RAPF are due to artificial noise introduced by the regularization step for the parameters and also to the hidden state estimation. In the EP-RAPF implementation we  have used 1,500 particles and the artificial noise is necessary to avoid degeneration of their weights. Note that within each time block the EP-RAFP is using the filtered states instead of the smoothed states. A combination of the RAFP with a MH step or a particle MCMC would improve the performance of the EP-RAFP algorithm in the estimation of both states and parameters, at a price of increasing computing . We leave the further study of this issue for a future communication.}

\begin{table}[p]
\begin{center}
\begin{scriptsize}
\begin{tabular}{c|ccccc|c|c}
\hline
\multicolumn{8}{c}{Parameter $\kappa$ MSE}\\
\hline
$J$ & \multicolumn{5}{c|}{EP-SMCMC}                                     & MCMC       &EP-RAPF\\
    & ($K=10$)      & ($K=20$)       & ($K=30$)  & ($K=40$)& ($K=50$)   &         &          \\
\hline
50  &   3.91       &    2.37       &     2.07    &   4.81  &  7.02       &  29.68 &  1.14     \\
100 &   2.09       &    1.68       &     2.11    &   3.06  &  6.75       &  32.33 &  0.90    \\
200 &   3.07       &    1.68       &     1.20    &   0.77  &  3.98       &  25.06 &  6.90    \\
\hline
\hline
\multicolumn{8}{c}{Parameter $\nu$ MSE}\\
\hline
$m$ & \multicolumn{5}{c|}{EP-SMCMC}                                     & MCMC          &EP-RAPF  \\
    & ($K=10$)      & ($K=20$)       & ($K=30$)  & ($K=40$)& ($K=50$)   &               &         \\
\hline
50  &  45.04       &    48.52      &    39.73    &   53.55 &  51.90  & 78.20  &        31.18        \\
100 &  28.41       &    29.14      &    20.52    &   39.01 &  43.52  & 56.19  &       101.14      \\
200 &  29.83       &    21.01      &    11.92    &   28.54 &  39.07  & 58.23  &        53.12      \\
\hline
\hline
\multicolumn{8}{c}{Parameter $\lambda$ MSE}\\
\hline
$m$ & \multicolumn{5}{c|}{EP-SMCMC}                                     & MCMC    &EP-RAPF  \\
    & ($K=10$)      & ($K=20$)       & ($K=30$)  & ($K=40$)& ($K=50$)   &         &         \\
\hline
50  &   0.02       &   0.01        &   0.01      &   0.01  &  0.01   &  0.01      &  0.02  \\
100 &   0.02       &   0.01        &   0.01      &   0.01  &  0.01   &  0.01      &  0.01  \\
200 &   0.01       &   0.01        &   0.01      &   0.01  &  0.01   &  0.01      &  0.01  \\
\hline
\end{tabular}
\end{scriptsize}
\end{center}
\caption{Mean square error for the parameters $\kappa$, $\nu$ and $\lambda$, for different size of the time blocks, $J$, and of the cross-sectional blocks, $K$, for the EP-SMCMC and a sequence of MCMC (1,500 iterations) started each time a block of observations is acquired and EP-RAPF with 1,500 particles. The average standard deviation, across algorithms and experiments, of the $\kappa$, $\nu$ and $\lambda$ MSE estimates is 0.3, 1.13 and 0.00001, respectively.}\label{TabMSE}
\end{table}

\begin{table}[p]
\begin{center}
\begin{tabular}{c|ccccc|c}
\hline
$J$    & \multicolumn{5}{c|}{EP-SMCMC}                                & MCMC      \\
       & ($r=10$)    & ($r=20$)      & ($r=30$)&  ($r=40$) &  ($r=50$)&           \\
\hline
50     &10.87        &6.66	         &2.66	   &1.64	   &1.01      & 49.39     \\
100    &5.13	     &2.45	         &1.88	   &0.70	   &0.47      & 28.10     \\
200    &2.57	     &1.43	         &0.71	   &0.32	   &0.19      & 15.27     \\
\hline
\end{tabular}
\end{center}
\caption{Computing time, in hours, for fixed cross-sectional block size $K=40$, for different size of the time blocks $J$ and different number of parallel CPU cores $r$, for the EP-SMCMC and a sequence of MCMC (1,500 iterations) started each time a block of observations is acquired.}\label{Tab}
\end{table}

Also, we compare EP-SMCMC, MCMC and EP-RAPF in terms of computing time. For the EP-SMCMC we consider  cross-sectional blocks of size $K=30$. The computing times are given in Table \ref{Tab} for different time acquisition rates $J$ (rows) and different CPUs $r$ working in parallel (columns). We conclude that the EP-SMCMC implemented on cluster multiprocessor system  can be up to 80 times faster than the standard MCMC. {These results are in line with the ones obtained in previous studies on parallel Monte Carlo methods (\cite{CGRD2013}, \cite{LeeYauGilDouHol13}, \cite{GD2012}). The calculation in  all our experiments have been carried out using double precision. If an application allows for a lower degree of precision, then single precision calculation can lead to large gains in computing time   as  documented by \cite{LeeYauGilDouHol13} in the context of GPU parallel computing.}

\subsection{Real data}
We consider a panel of 12,933 assets of the US stock market and collect prices at a daily frequency from 29 December 2000 to 22 August 2014, which yields 3,562 time observations. Then we compute logarithmic returns for all stocks and obtain a dataset of 46,067,346 observations. 

\begin{figure}[t]
\begin{center}
\begin{tabular}{c}
\vspace{-150pt}\\
\includegraphics[width=0.9\textwidth, height=400pt]{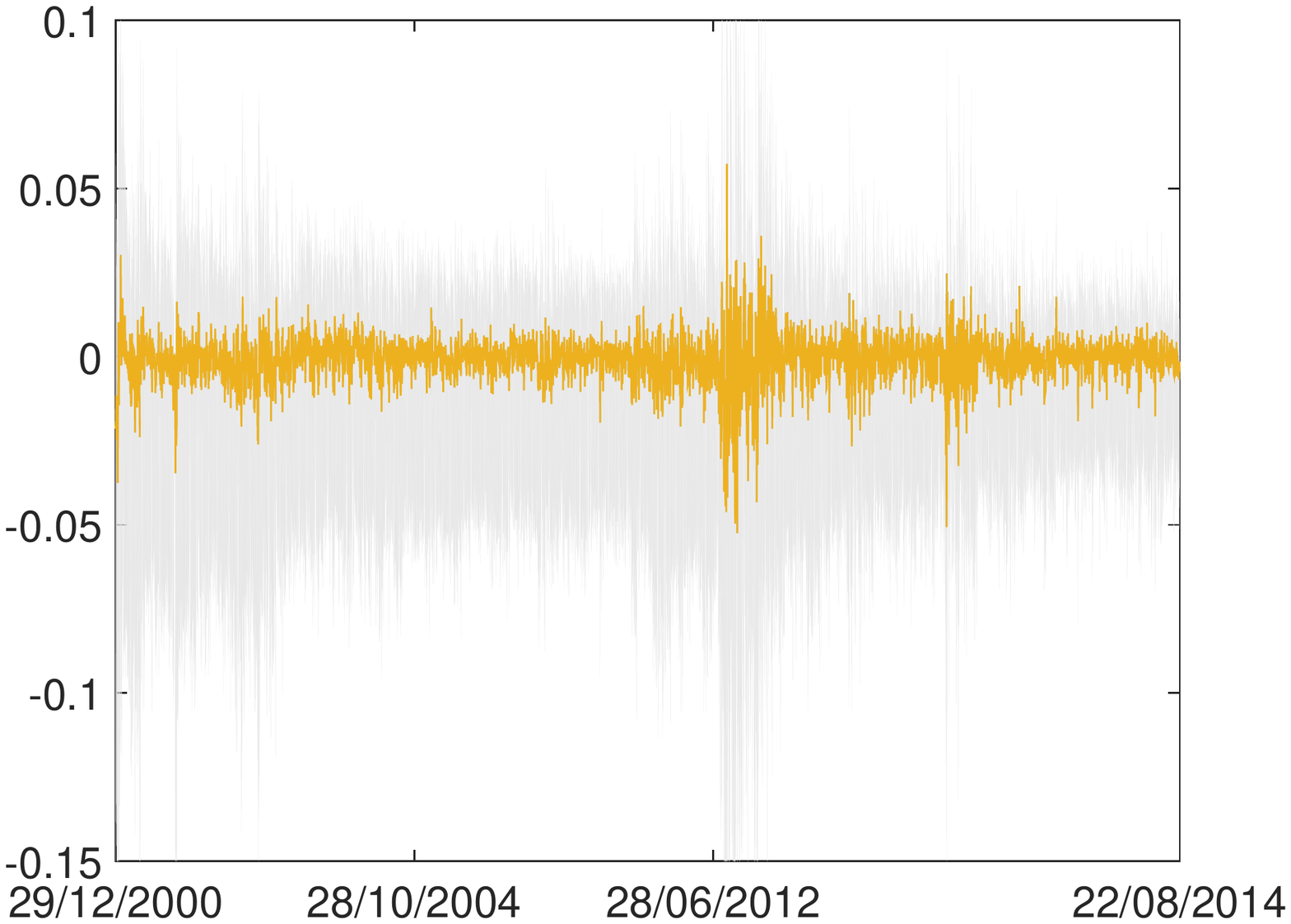}\\
\vspace{-150pt}\\
\includegraphics[width=0.9\textwidth, height=400pt]{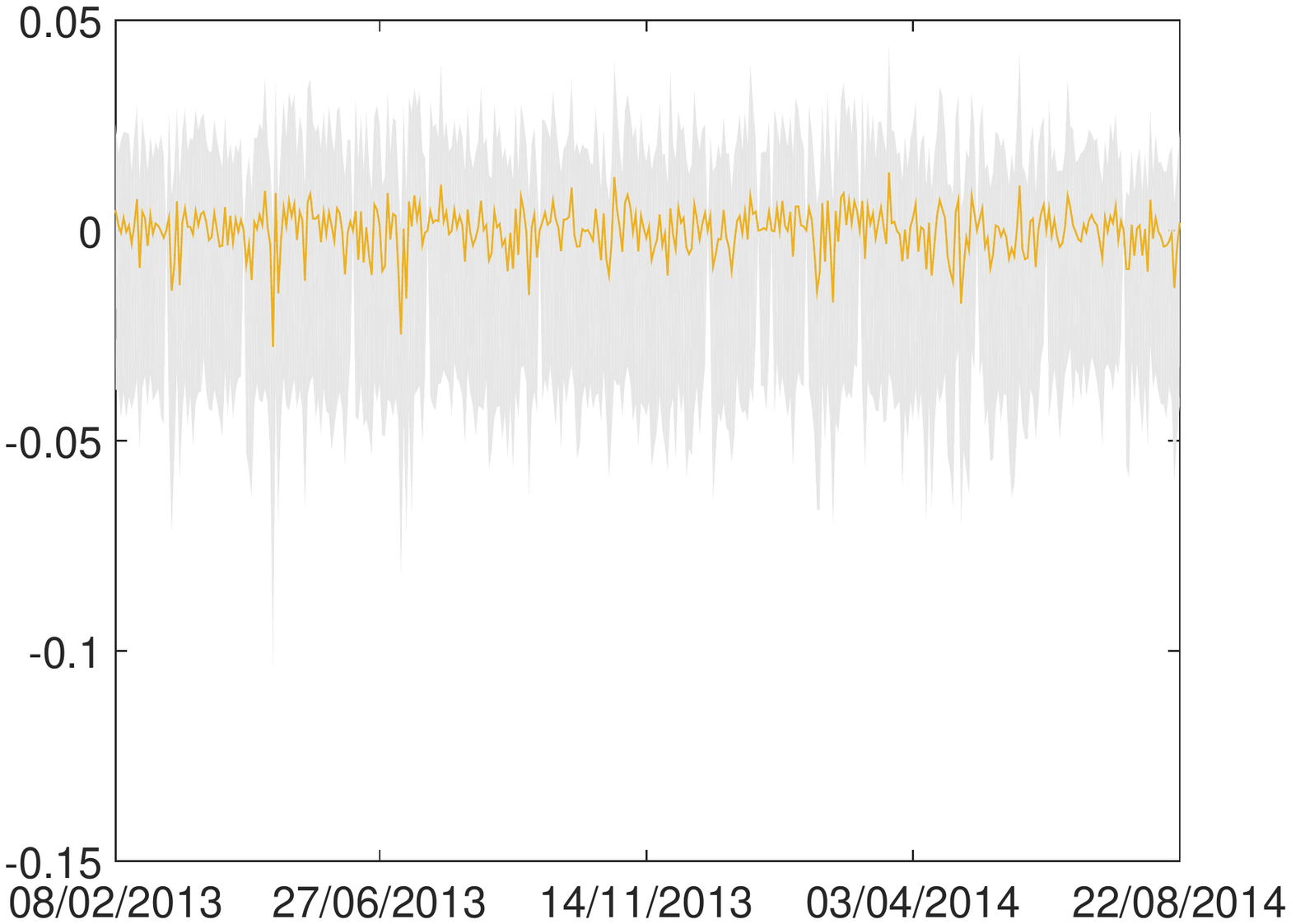}
\end{tabular}
\end{center}
\caption{Quantiles at the 5\% and 90\% (gray area) and mean (solid line) of the cross-sectional daily log-return distribution. Returns for all the 12,933 assets (top panel) of the US stock market, from 29 December 2000 to 22 August 2014 and for a subsample of 6,799 assets (bottom panel) from 8 February to 22 August 2014 for.}\label{Data}
\end{figure}

In order to control for the liquidity of the assets and consequently for long sequences of zero returns, we impose that each stock has been traded a number of days corresponding to at least 40\% of the sample size. Also we focus on the last part of the sample which include observations from 8 February 2013 to 22 August 2014. After cleaning the dataset we obtain 6,799 series and 400 time observations, and the size of the dataset reduces to 2,846,038. The original and the cleaned datasets are give in Fig. \ref{Data}.

In order to capture the time variations in the volatility of the series we consider the panel state space model  presented in the previous section. We apply our EP-SMCMC sampler to the panel of 6,799 time series. The data are acquired sequentially over time in blocks of  100 time observations and each panel consists of  $K=40$ series.  At each point in time, for each parameter $k$, $\nu$ and $\lambda$ we obtain 500 posterior densities from each parallel SMCMC chain (see \ref{Seq1}-\ref{Seq3} in Appendix \ref{appSMCMC} an example for $t=100,200,300,400$).

\begin{figure}[p]
\begin{center}
\vspace{-20pt}
\begin{tabular}{c}
\includegraphics[width=0.86\textwidth, height=275pt]{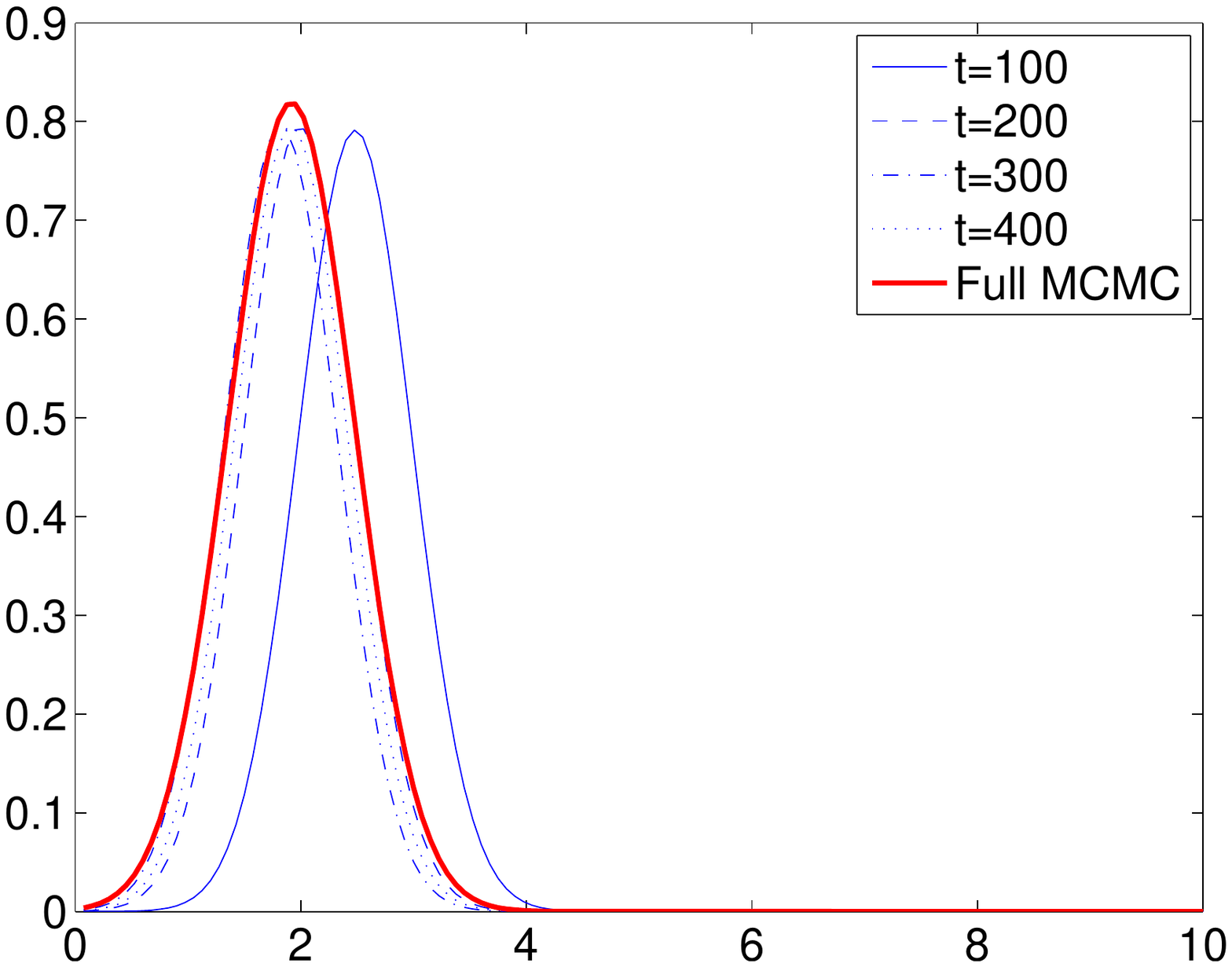}\\
\vspace{-150pt}\\
\includegraphics[width=0.86\textwidth, height=275pt]{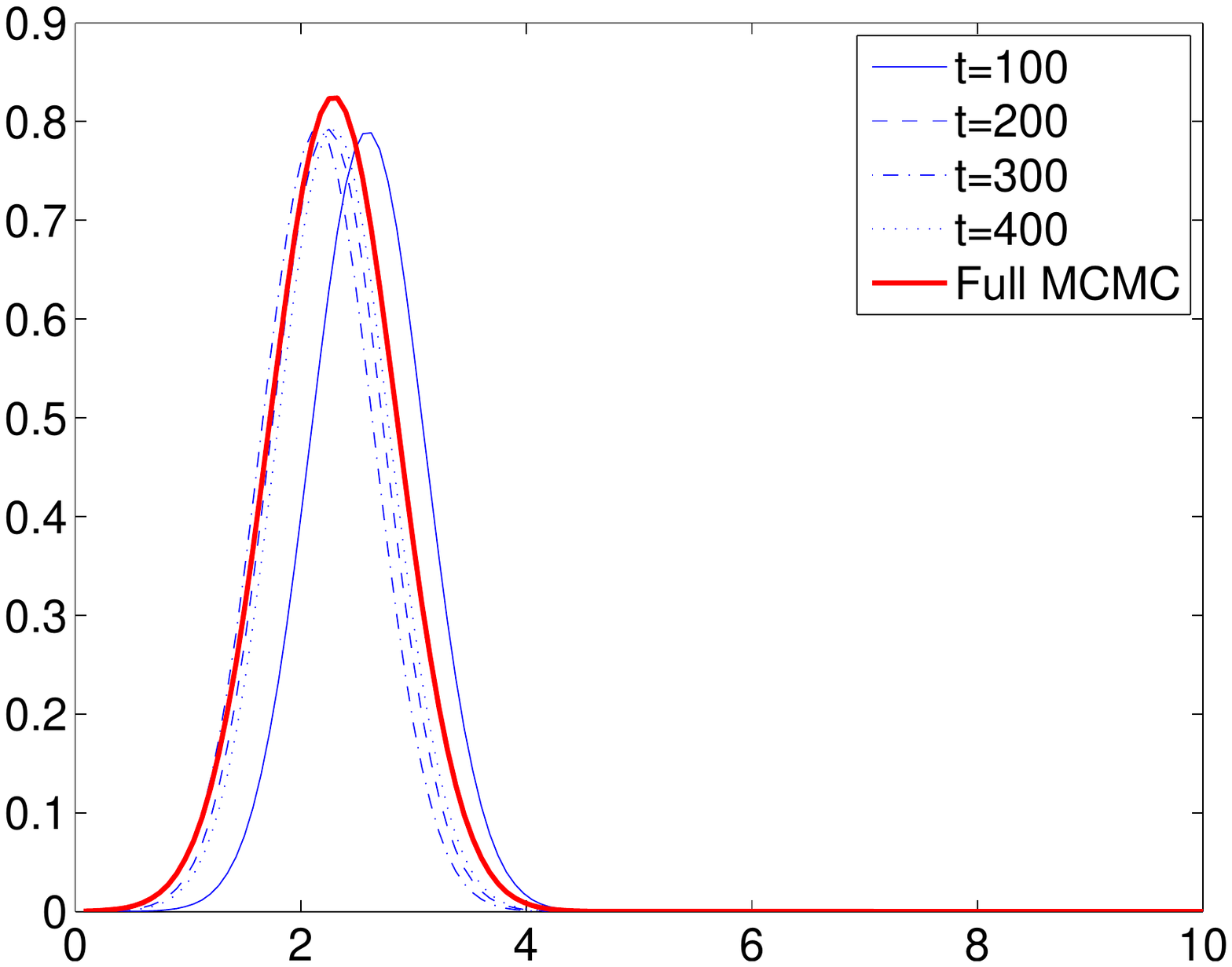}\\
\vspace{-150pt}\\
\includegraphics[width=0.86\textwidth, height=275pt]{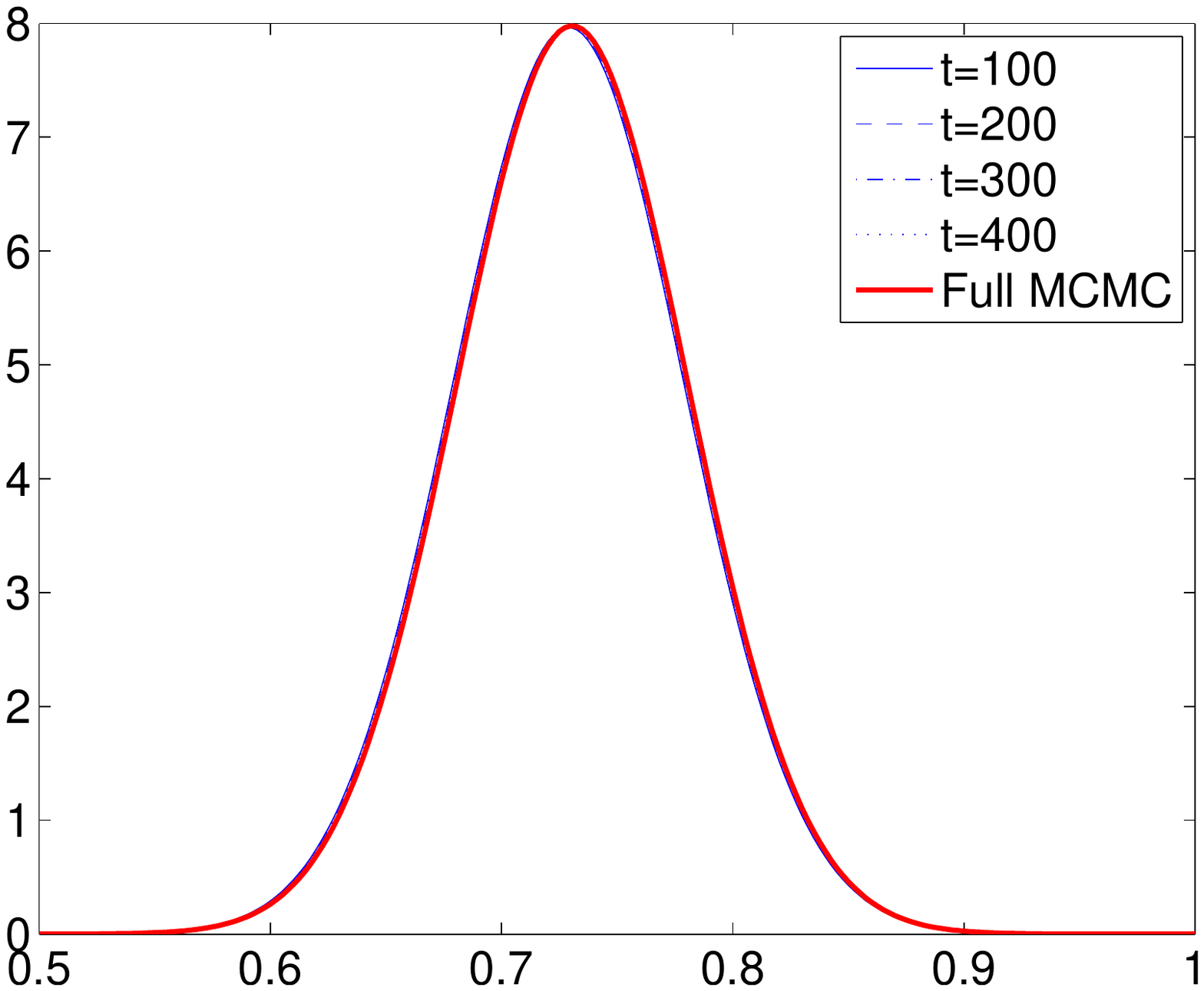}
\end{tabular}
\end{center}
\vspace{-10pt}
\caption{Sequential estimation of the posterior densities of the parameter $k$ (top panel), $\nu$ (middle panel) and $\lambda$  (bottom panel) at different points in time $t$. The whole sample and pooled data posterior densities of the parameter are given by the dashed line.}\label{SeqEmbarrassing}
\end{figure}

Figure \ref{SeqEmbarrassing} shows the sequential posterior inference after the merge step of the embarrassingly parallel SMCMC algorithm is applied to the output of the SMCMC samples. At each point in time we obtain an approximated posterior density for the whole cross-section from the embarrassingly parallel step (see solid lines in \ref{SeqEmbarrassing} for $t=100,200,300,400$).

{The approximation of the posterior produced by the EP-SMCMC is close to the approximation based on a  MCMC analysis of the full posterior. This can be seen in Fig. \ref{SeqEmbarrassing}) where the solid line shows the EP-SMCMC approximation at the last update ($t=400$) and the dashed line represents  the  full-sample estimate based on a standard MCMC with 25,000 iterations after a burn-in period of 20,000 iterations. The posterior mean and posterior quantiles approximated by EP-SMCMC are given in Tab. \ref{TabPostEst}.  In order to approximate the high posterior density (HPD) intervals and the posterior mean we generate samples from the posterior distribution  given in Eq. \ref{joint-posterior} by applying the independent Metropolis within Gibbs 
algorithm given in \cite{NeiWanXin14}.
\begin{table}{t}
\begin{center}
\begin{tabular}{c|cc|cc}
\hline
                      &\multicolumn{2}{|c|}{EP-SMCMC}&\multicolumn{2}{|c}{MCMC}\\
 $\theta$             & $\hat{\theta}$ & HPD  & $\hat{\theta}$   & HPD \\
\hline
$\kappa$              &   0.66       &   (0.61,0.71)           &   0.54         &(0.31,0.58)  \\ 
$\nu$                 &   0.87       &   (0.79,0.95)           &   0.91         &(0.86,1.01)  \\ 
$\lambda$             &   0.98       &   (0.97,0.99)           &   0.96         &(0.94,0.97)  \\ 
\hline 
\end{tabular}
\end{center}
\caption{EP-SMCMC and MCMC approximation of the posterior mean ($\hat{\theta}$) and  the 95\% high probability density region HPD given by the 2.5\% and 97.5\% inter-quantile inveral $(q_{0.025},q_{0.975})$.}\label{TabPostEst}
\end{table}
}

\section{Conclusion}\label{sec:conclusion}
We propose a new MCMC algorithm which combines embarrassingly parallel MCMC and Sequential MCMC. The algorithm is developed for data that exhibit dependent patterns, in particular for large sets of time series for which a standard MCMC-based analysis would be very slow. {Here we take advantage of the independence between the unit-specific observations and latent variables of the panel to partition the data and factor the full posterior in a product of partial posteriors. In the absence of clear independent panel units, an interesting and difficult question concerns alternative strategies to divide the data and combine the partial posterior samples.} 
 
It is apparent that the development of novel MCMC algorithms for big data is evolving rapidly. While ``divide and conquer'' strategies continue to develop, one must devise techniques to handle the additional approximations that are introduced by the current existing methods, including EP-SMCMC.  Quantifying and controlling the error introduced by these approximations remains central to the success of MCMC for Big Data.

\section*{Acknowledgement}
We thank the Editor, the Associate Editor and three referees for comments and suggestions that have greatly improved the paper.
This research used the SCSCF multiprocessor cluster system at University Ca' Foscari of Venice. Roberto Casarin's research is supported by funding from the European Union, Seventh Framework Programme FP7/2007-2013 under grant agreement SYRTO-SSH-2012-320270, by the Institut Europlace of  Finance, ``Systemic Risk grant'', the Global Risk Institute in Financial Services, the Louis Bachelier Institute, ``Systemic Risk Research Initiative'', and by the Italian Ministry of Education, University and Research (MIUR) PRIN 2010-11 grant MISURA. Radu Craiu's research is supported by the Natural Sciences and Engineering Research Council of Canada. Fabrizio Leisen's research is supported by the European Community's Seventh Framework Programme FP7$/$2007-2013 under grant agreement number 630677.

\newpage

\clearpage

\renewcommand{\thesection}{A}
\renewcommand{\theequation}{A.\arabic{equation}}
\renewcommand{\thefigure}{A.\arabic{figure}}
\renewcommand{\thetable}{A.\arabic{table}}
\setcounter{table}{0}
\setcounter{figure}{0}
\setcounter{equation}{0}

\appendix
\section{Computational details}\label{app}
\subsection{Transition kernel}
As regards the Step 1.1. of the transition kernel, the distribution $\pi(\boldsymbol{\theta}|X_{it}^{(l,j)},Y_{it})\propto \pi(\boldsymbol{\theta},X_{it}^{l,j}|Y_{it})$ is not tractable and we applied a Metropolis-Hastings. At the $j$-th iteration of the $l$-th SMCMC chain we generate the MH proposal from a Gaussian random walk on the transformed parameter space $\tilde{\theta}_{1}=\log(\kappa)$, $\tilde{\theta}_{2}=\log(\nu)$ and $\tilde{\theta}_{3}=\log((1-\lambda)/\lambda)$.

In the Step 1.2 of the transition kernel, we exploit the tractability of the state space model and apply a multi-move Gibbs sampler, where the hidden states $\mathbf{x}_{i1:t}$ are updated in one step. By applying Proposition 1 in \cite{WinCar14} with $m=1$, one gets the following filtered, and prediction distributions
\begin{eqnarray}
x_{it}|\boldsymbol{\theta},Y_{it}&\sim&\mathcal{G}a((\kappa+k)/2,\kappa \sigma^{2}_{it}/2)\\
x_{it+1}|\boldsymbol{\theta},Y_{it}&\sim&\mathcal{G}a(\kappa/2,\kappa\sigma^{2}_{it}/2/\lambda)\end{eqnarray}
where $\sigma_{it}^{2}=y_{it}+\lambda \sigma_{it-1}^{2}$, and the backward smoothed distribution
\begin{equation}
x_{it}|\boldsymbol{\theta},x_{it+1},Y_{it}\sim\lambda x_{it+1}+z_{it+1},\quad z_{it+1}\sim \mathcal{G}a(\kappa/2,(\kappa\sigma_{it}^{2})/2).
\end{equation}
which is used to generate $X_{it}^{(l,j)}$ given $\boldsymbol{\theta}^{(l,j)}$ and $Y_{it}$.

\subsection{Jumping kernel}
As regard to the jumping kernel, when $d_{t+1}=d_{t}+1$, it is given by a Gibbs sampler transition kernel with full conditional distribution
\begin{equation}
x_{it+1}|\boldsymbol{\theta},Y_{i,t+1}\sim\mathcal{G}a((\kappa+k)/2,\kappa \sigma^{2}_{it+1}/2)
\end{equation}

\renewcommand{\thesection}{B}
\renewcommand{\theequation}{B.\arabic{equation}}
\renewcommand{\thefigure}{B.\arabic{figure}}
\renewcommand{\thetable}{B.\arabic{table}}
\setcounter{table}{0}
\setcounter{figure}{0}
\setcounter{equation}{0}

\newpage

\clearpage

\section{Computational details for the EP-RAPF}\label{AppSMC}
Let us consider the following reparametrization $\boldsymbol{\theta}=(\kappa, \log(\nu-\kappa+1), \log((1+\lambda)/(1-\lambda))$. Given the initial sets of weighted random samples $\left\{\mathbf{x}^j_{it_0},\boldsymbol{\theta}^{j}_{it_0},w^j_{it_0}\right\}_{j=1}^{N}$, $i=1,\ldots,M$ the regularized APF performs the following steps, for $t_0< t\leq T-1$, $i=1,\ldots,M$ and $j=1,\ldots,N$:

\begin{itemize}
\item[\textit{(i)}] Simulate $r^j_i\sim q(r_i)\propto\sum_{l=1}^{N}w_{it}^{l}\delta_{l}(r_i)$ where
\small{
$$
w_{it}^{l}\propto w_{it-1}^l \prod_{k=k_{i-1}+1}^{k_{i}}\mathcal{G}a(y_{kt+1}|\kappa_{it}^{r_i}/2,\kappa^{r_i}_{it}\mu_{kt+1}^{r_i} /2)
$$
}
with 
$$
\mu_{kt+1}^{r}= \frac{x_{kt}^r}{\lambda^{r}_{it}}\frac{\nu^{r}_{it}}{\nu^{r}_{it}+\kappa^{r}_{it}}
$$.

\item[\textit{(ii)}] Simulate
$\boldsymbol{\theta}^{j}_{it+1}$ $\sim$ $\mathcal{N}\left(a\boldsymbol{\theta}^{r^j_i}_{it}+(1-a)\bar{\boldsymbol{\theta}}_{it},h^{2}V_{it}\right)$ where $V_{it}$ and $\bar{\boldsymbol{\theta}}_{it}$ are the empirical variance matrix and the empirical mean respectively and $a\in[0,1]$ and
$h^{2}=(1-a^{2})$,

\item[\textit{(iii)}] Simulate $x_{kt+1}^j\sim
x_{kt}^{r^j_i}\psi_{it}^{j}/\lambda^j_{it+1}$, with $\psi_{kt}^{j}\sim\mathcal{B}e(\nu_{it}^{j}/2,\kappa_{it}^{j}/2)$ for $k=k_{i-1}+1,\ldots,k_i$

\item[\textit{(iv)}] Update the weights
\small{
$$
\omega_{it+1}^j\propto\prod_{k=k_{i-1}+1}^{k_{i}}\frac{\mathcal{G}a(y_{kt+1}|\kappa_{it}^{j}/2,\kappa_{it}^{j}x_{kt+1}^j/2)}{\mathcal{G}a(y_{kt+1}|\kappa_{it}^{r_i^j}/2,\kappa_{it}^{r_i^j}\mu_{kt+1}^{r_i^j}/2)}
$$
}
\item[\textit{(v)}] If $\mbox{ESS}_{t+1}<\varepsilon$, simulate $\left\{\mathbf{x}^j_{it+1},\boldsymbol{\theta}^{j}_{it+1}\right\}_{j=1}^{N}$ from $\left\{\mathbf{x}^j_{it+1},\boldsymbol{\theta}^{j}_{it+1},\omega^j_{it+1}\right\}_{j=1}^{N}$ and set $w^j_{it+1}=1/N$. Otherwise set $w^j_{it+1}=\omega^{j}_{it+1}$
\end{itemize}
where 
\small{
$$
\mbox{ESS}_t=\frac{N}{\displaystyle 1+N\sum_{i=1}^N\left(\omega_t^i-N^{-1}\sum_{i=1}^N\omega_t^i\right)^2 \bigg/ \left(\sum_{i=1}^N\omega_t^i\right)^2 }\,.
$$
}
is the effective sample size.

In the merge step of our EP-RAPF, the particle set $\left\{\boldsymbol{\theta}^{j}_{it},\omega^j_{it}\right\}_{j=1}^{N}$ is used to build the following approximation of the posterior distribution 
\begin{equation}
\pi_{t}(\boldsymbol{\theta})=\prod_{i=1}^{M}\hat{\pi}_{it}(\boldsymbol{\theta})
\end{equation}
by applying the embarrassingly parallel algorithm, as in the EP-SMCMC.

\renewcommand{\thesection}{C}
\renewcommand{\theequation}{C.\arabic{equation}}
\renewcommand{\thefigure}{C.\arabic{figure}}
\renewcommand{\thetable}{C.\arabic{table}}
\setcounter{table}{0}
\setcounter{figure}{0}
\setcounter{equation}{0}

\newpage

\clearpage

\section{SMCMC output}\label{appSMCMC}
\begin{figure}[p]
\begin{center}
\begin{tabular}{cc}
$t=100$ & $t=200$ \\
\includegraphics[width=0.5\textwidth, height=180pt]{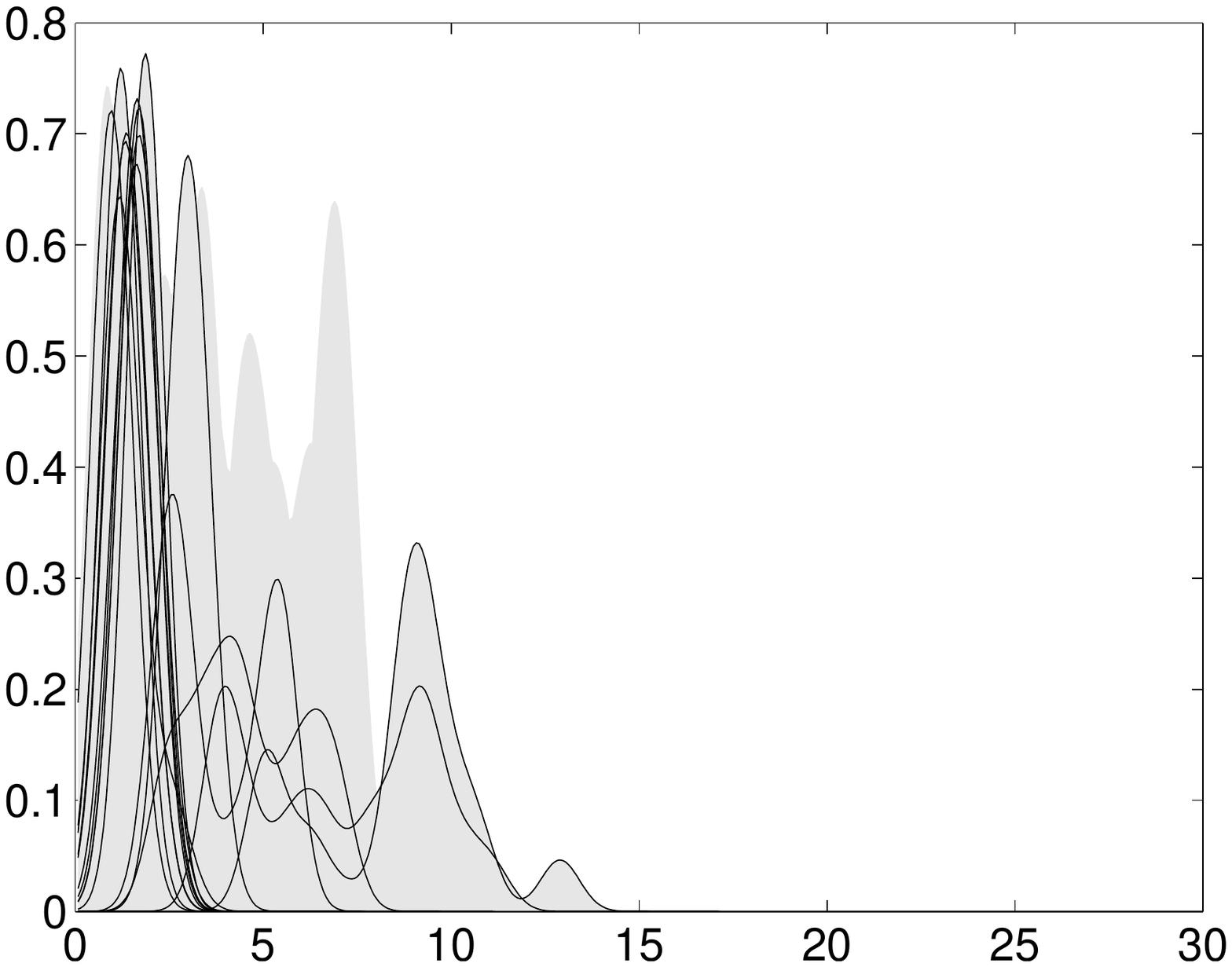}
&
\includegraphics[width=0.5\textwidth, height=180pt]{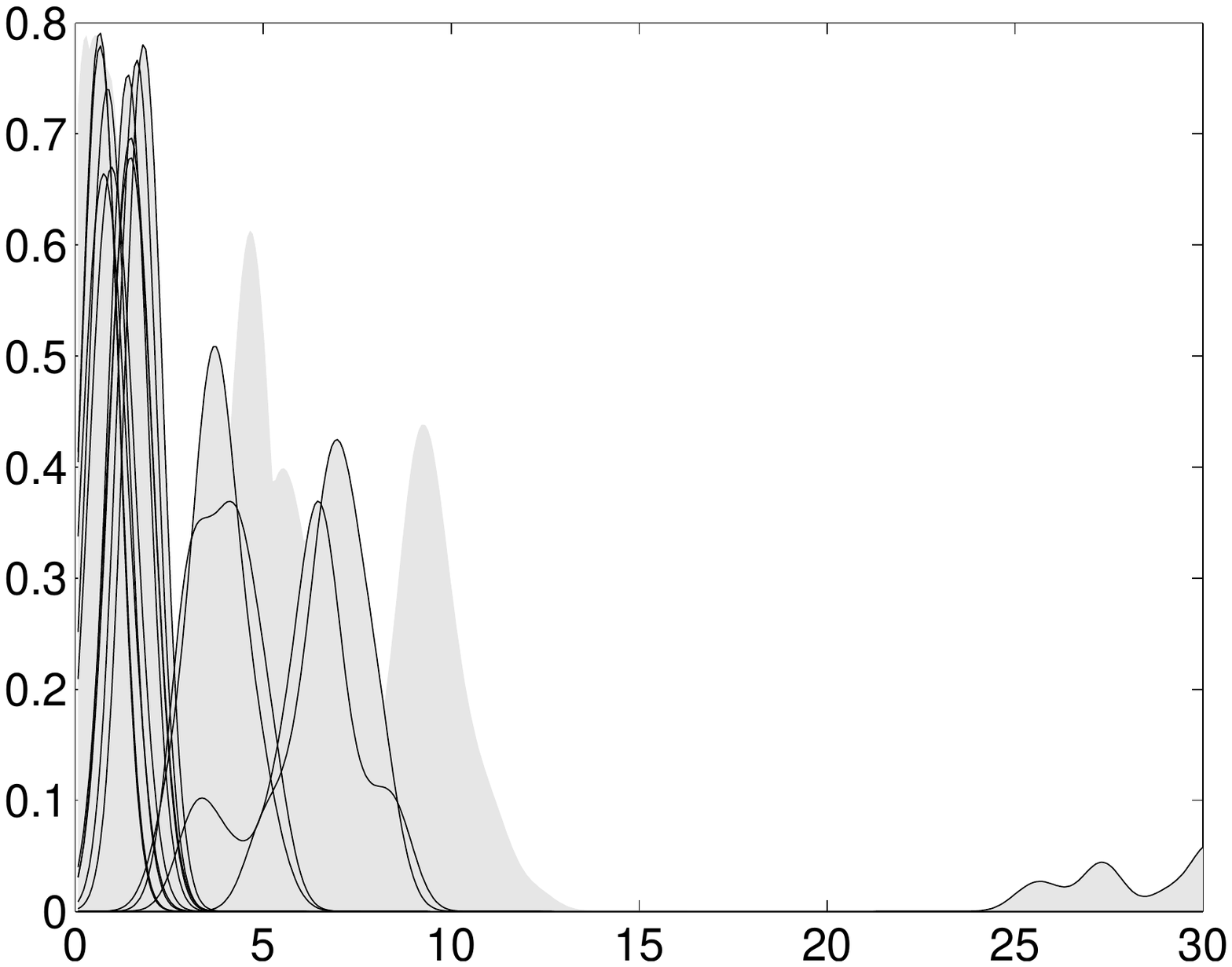}\\
$t=300$ & $t=400$\\
\includegraphics[width=0.5\textwidth, height=180pt]{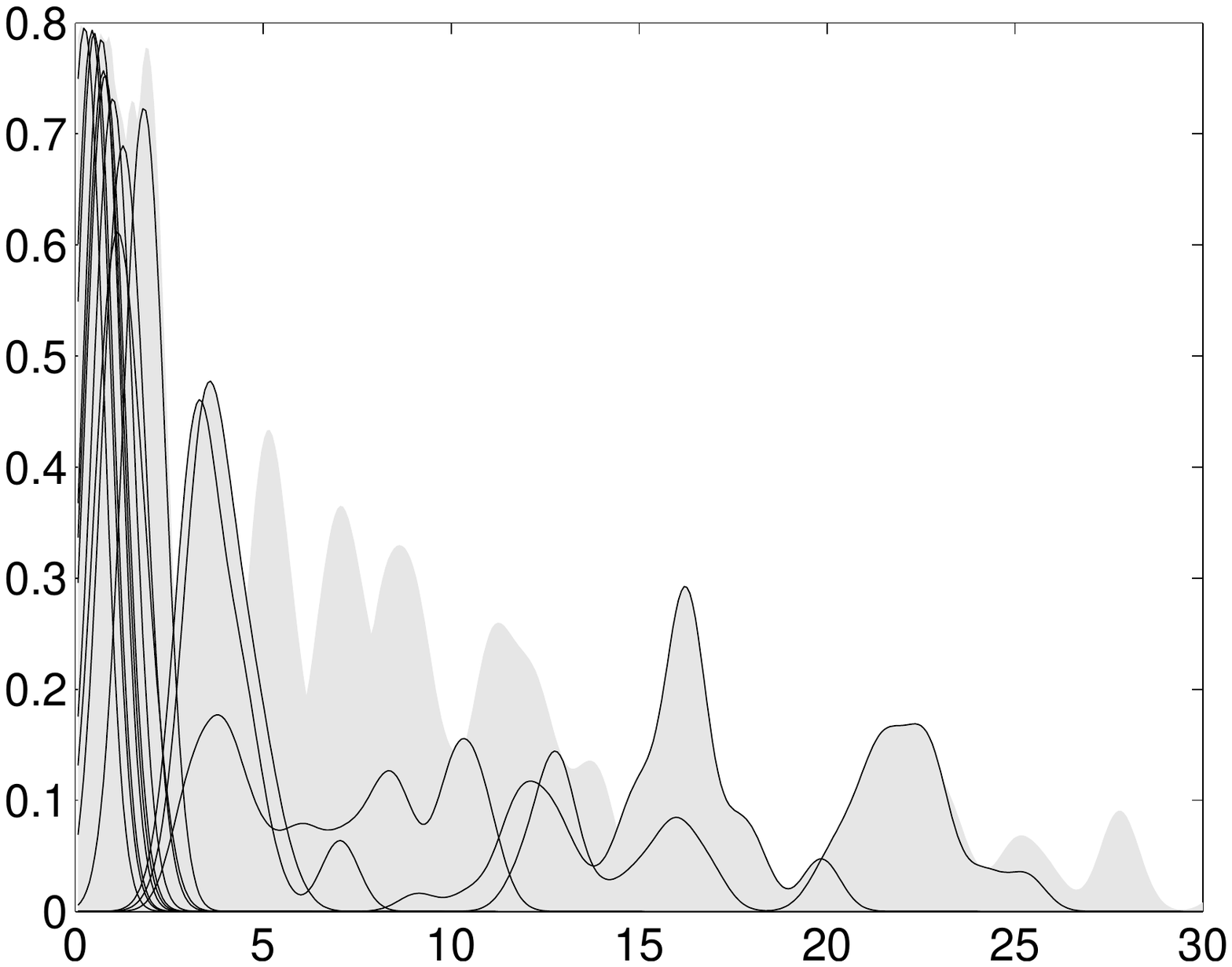}
&
\includegraphics[width=0.5\textwidth, height=180pt]{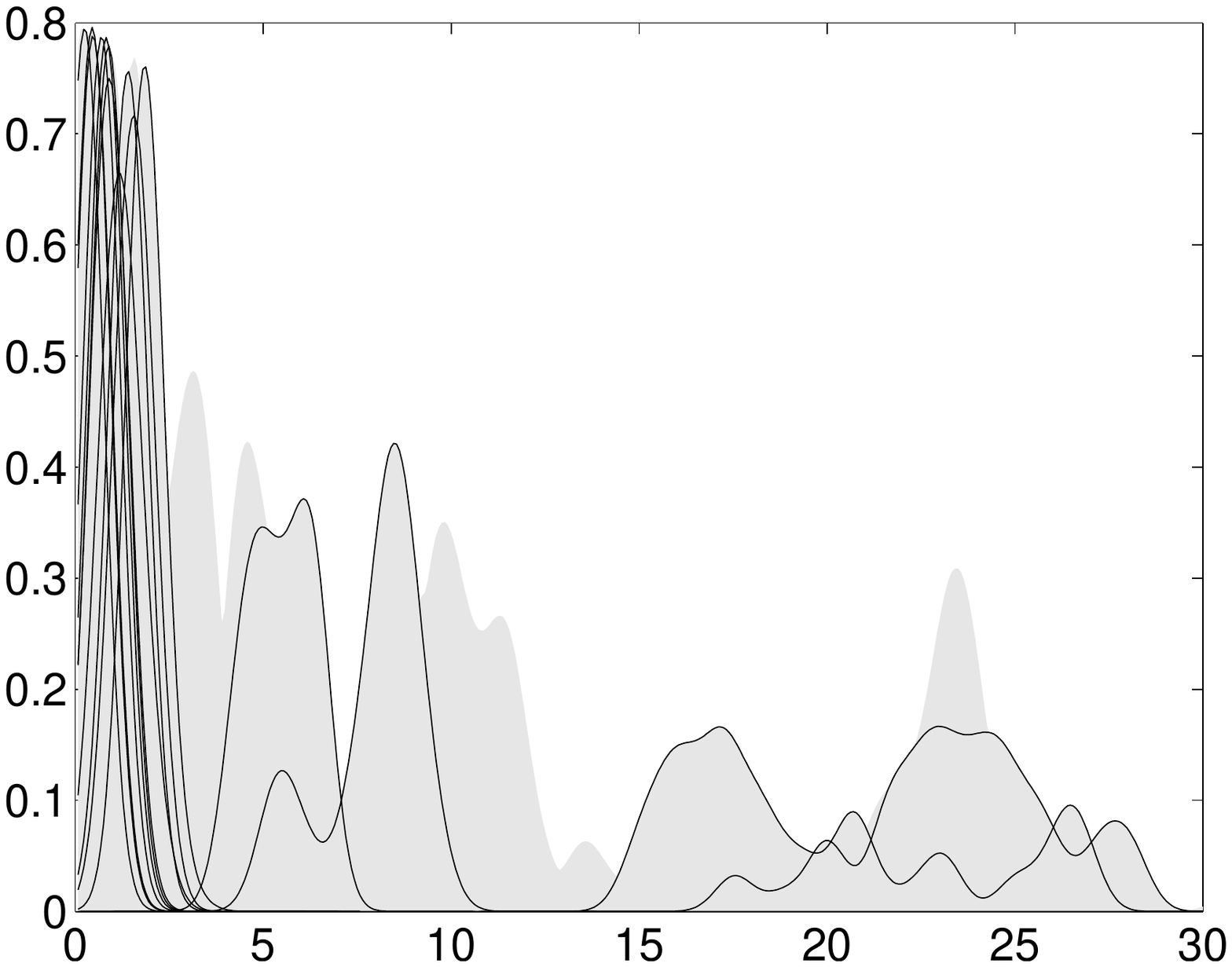}\\
\end{tabular}
\end{center}
\caption{Sequential estimation of the posterior densities of the parameter $k$, for the different blocks of observations (different lines) and at different point in time $t$ (different plots). For expository purposes the $\lfloor m/K \rfloor=169$ lines have been subsampled and the grey area represents the area below the envelope of the $N$ densities.}\label{Seq1}
\end{figure}

\begin{figure}[p]
\begin{center}
\begin{tabular}{cc}
$t=100$ & $t=200$ \\
\includegraphics[width=0.5\textwidth, height=180pt]{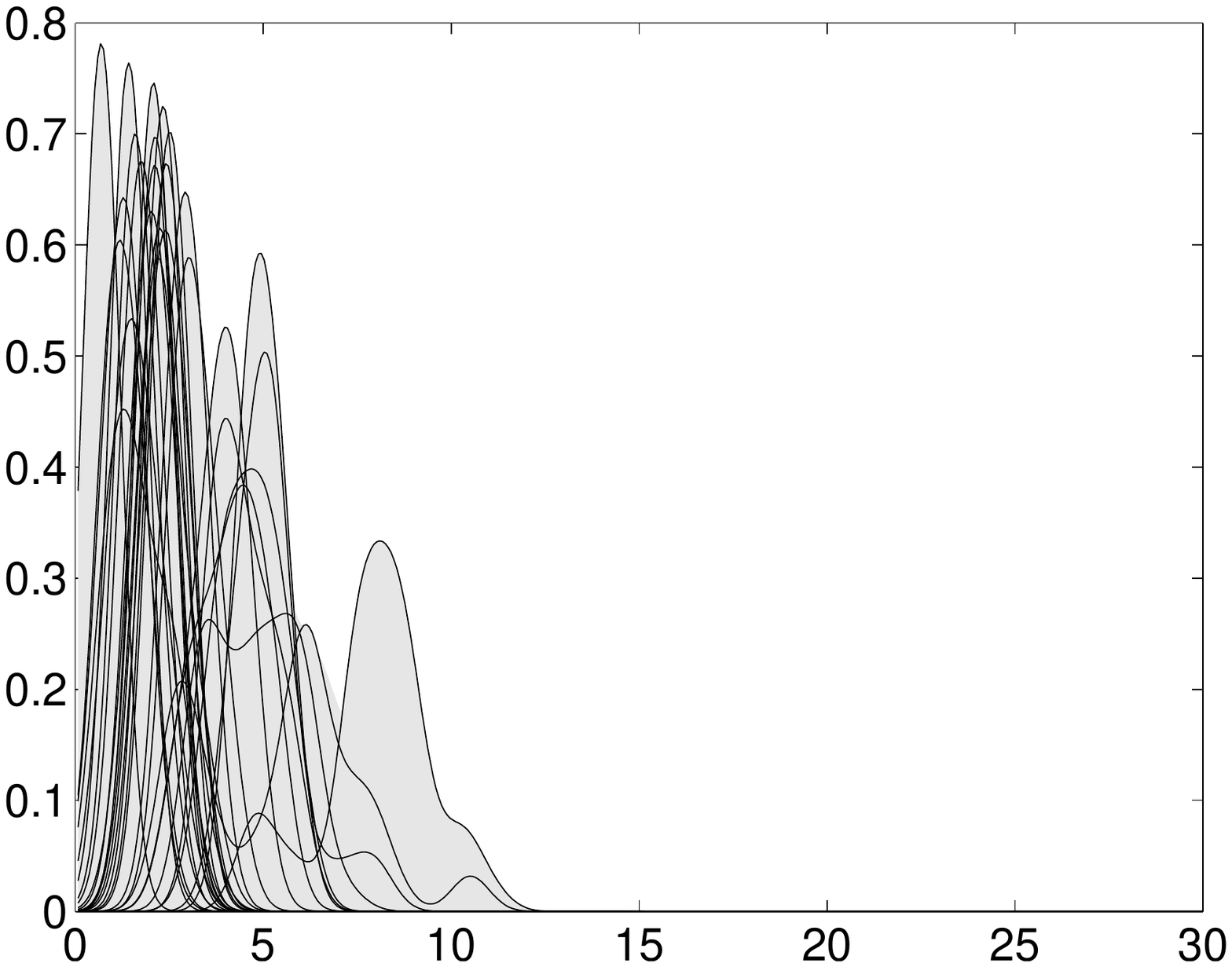}
&
\includegraphics[width=0.5\textwidth, height=180pt]{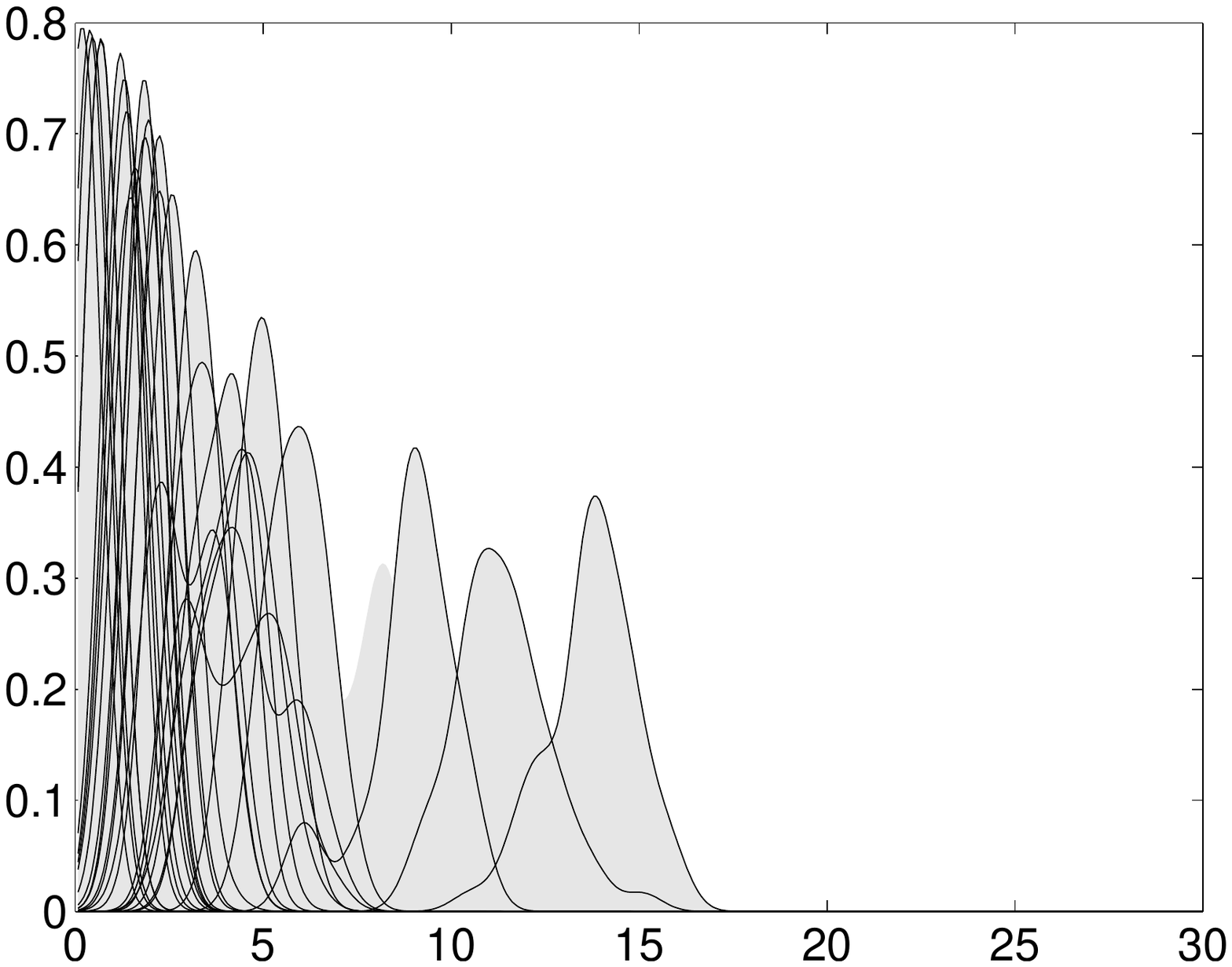}\\
$t=300$ &$t=400$\\
\includegraphics[width=0.5\textwidth, height=180pt]{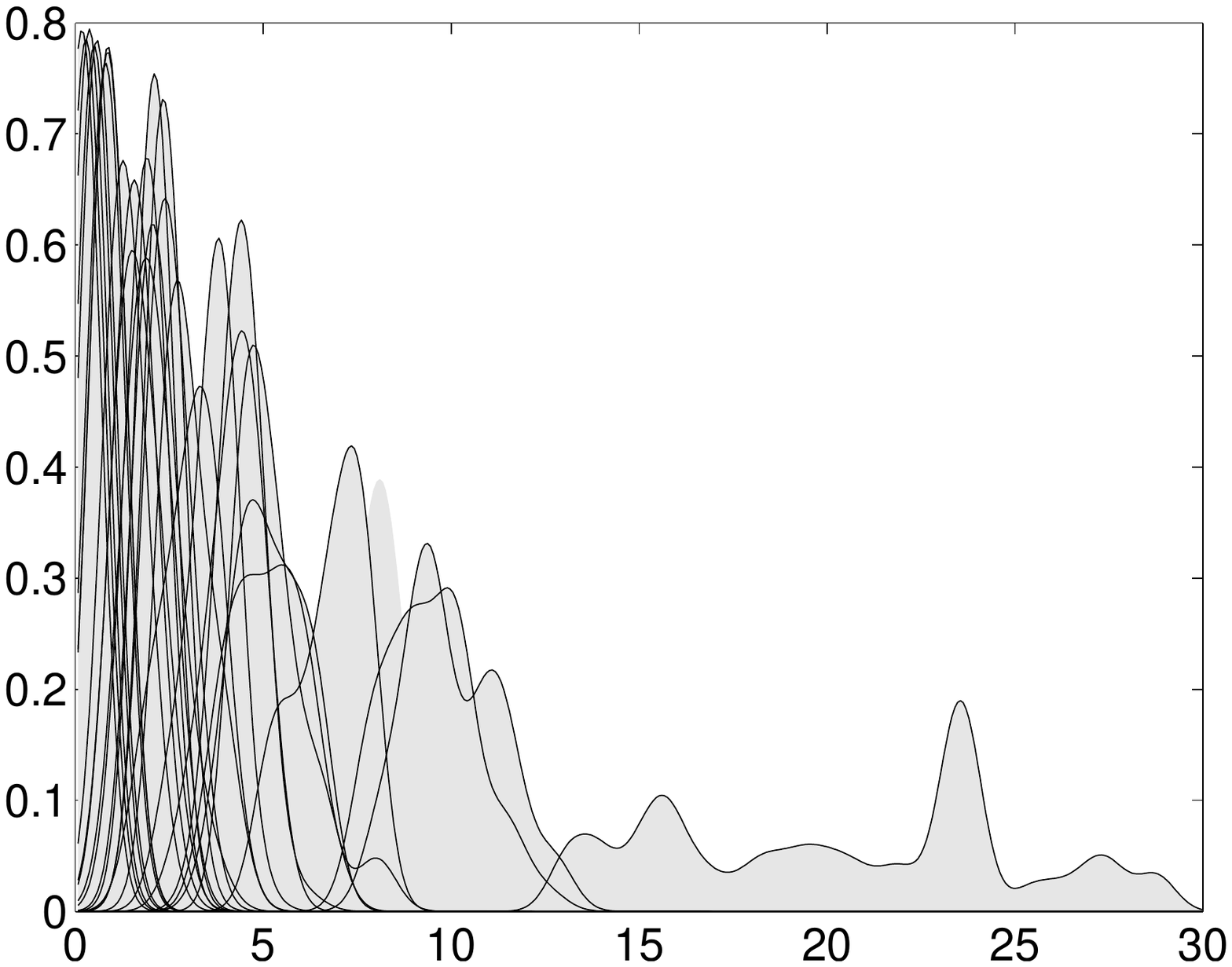}
&
\includegraphics[width=0.5\textwidth, height=180pt]{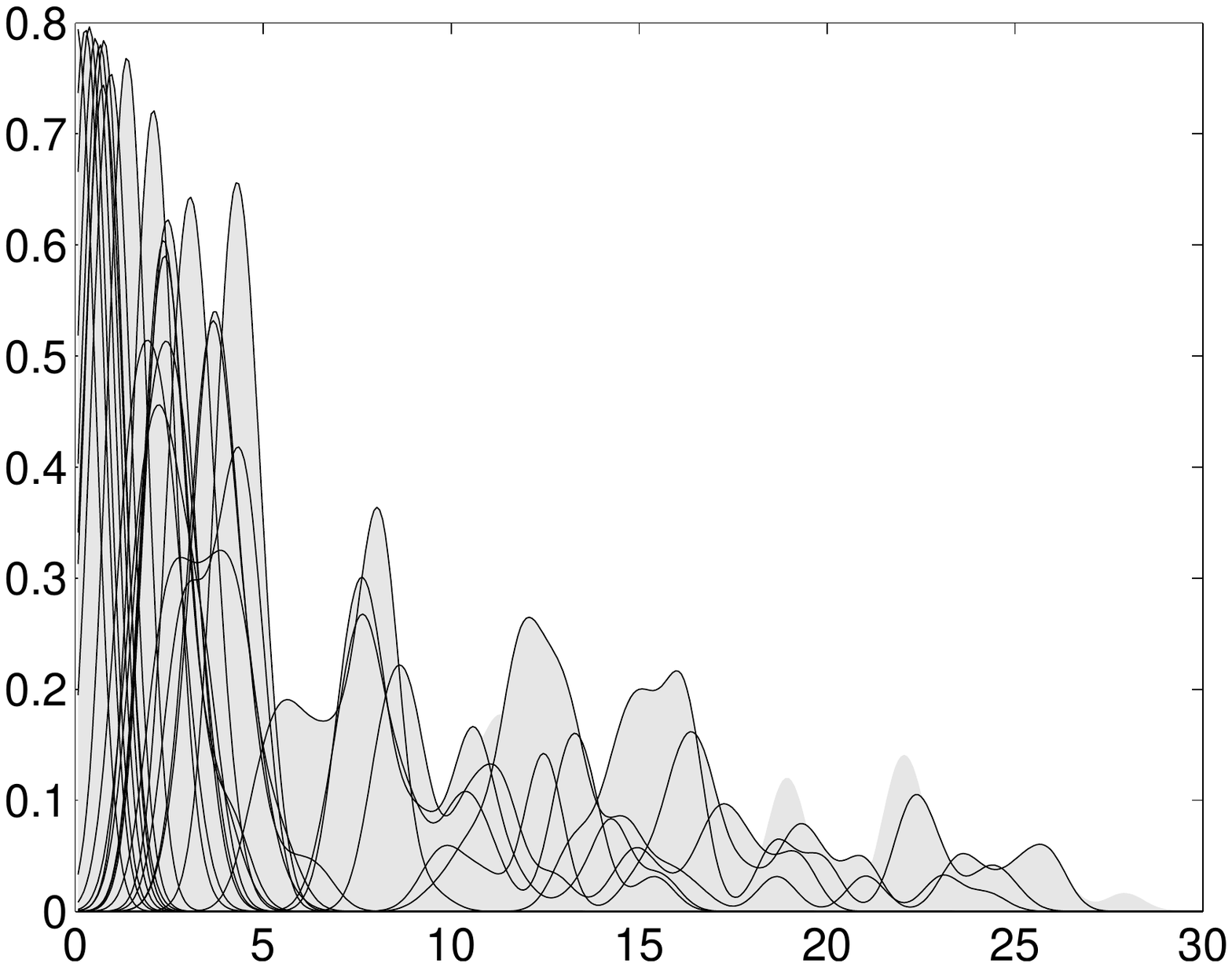}\\
\end{tabular}
\end{center}
\caption{Sequential estimation of the posterior densities of the parameter $\nu$, for the different blocks of observations  and at different point in time $t$. For expository purposes the $\lfloor m/K \rfloor=169$ lines have been subsampled and the grey area represents the area below the envelope of the $N$ densities.}\label{Seq2}
\end{figure}

\begin{figure}[p]
\begin{center}
\begin{tabular}{cc}
$t=100$ & $t=200$\\
\includegraphics[width=0.5\textwidth, height=180pt]{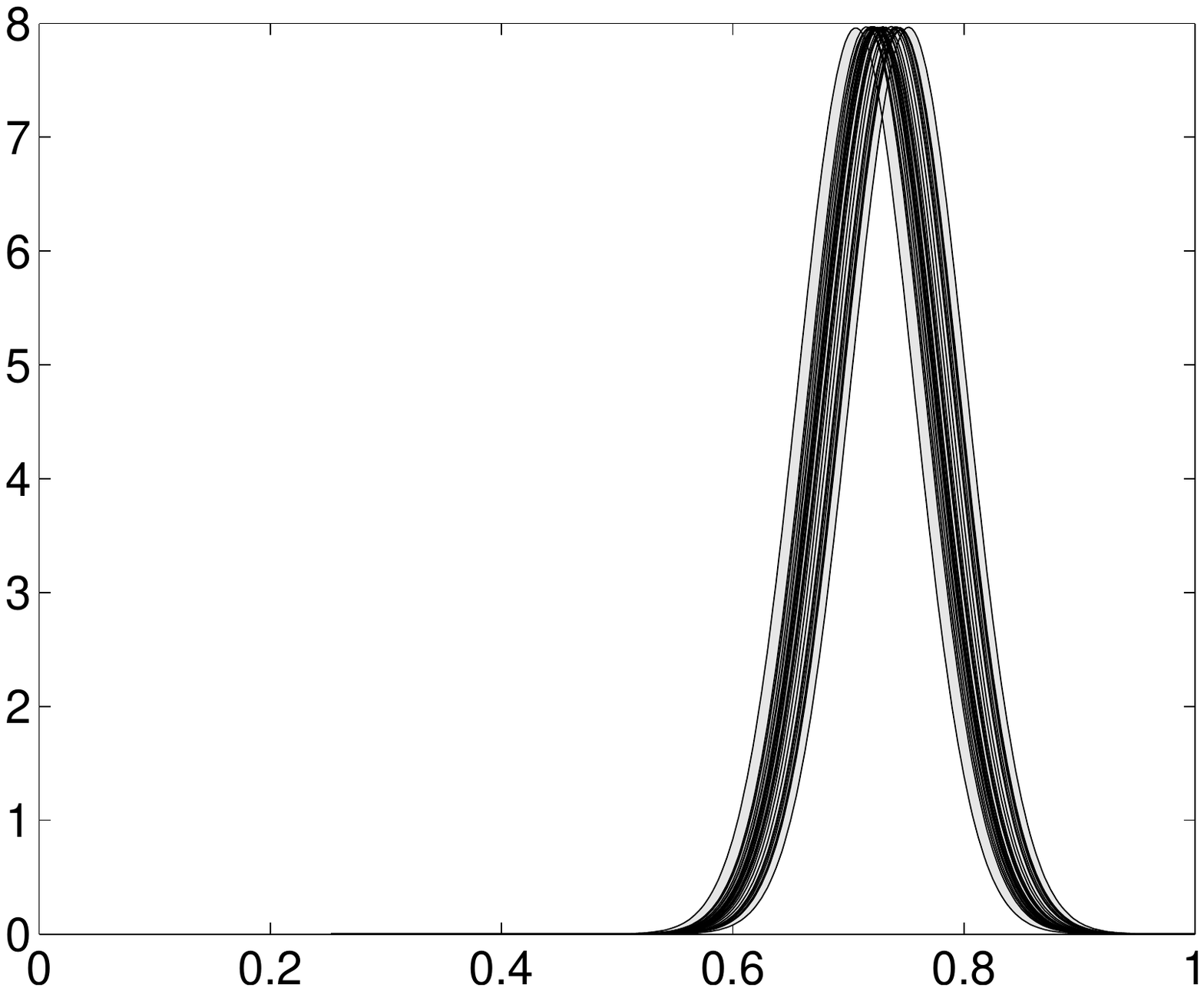}
& 
\includegraphics[width=0.5\textwidth, height=180pt]{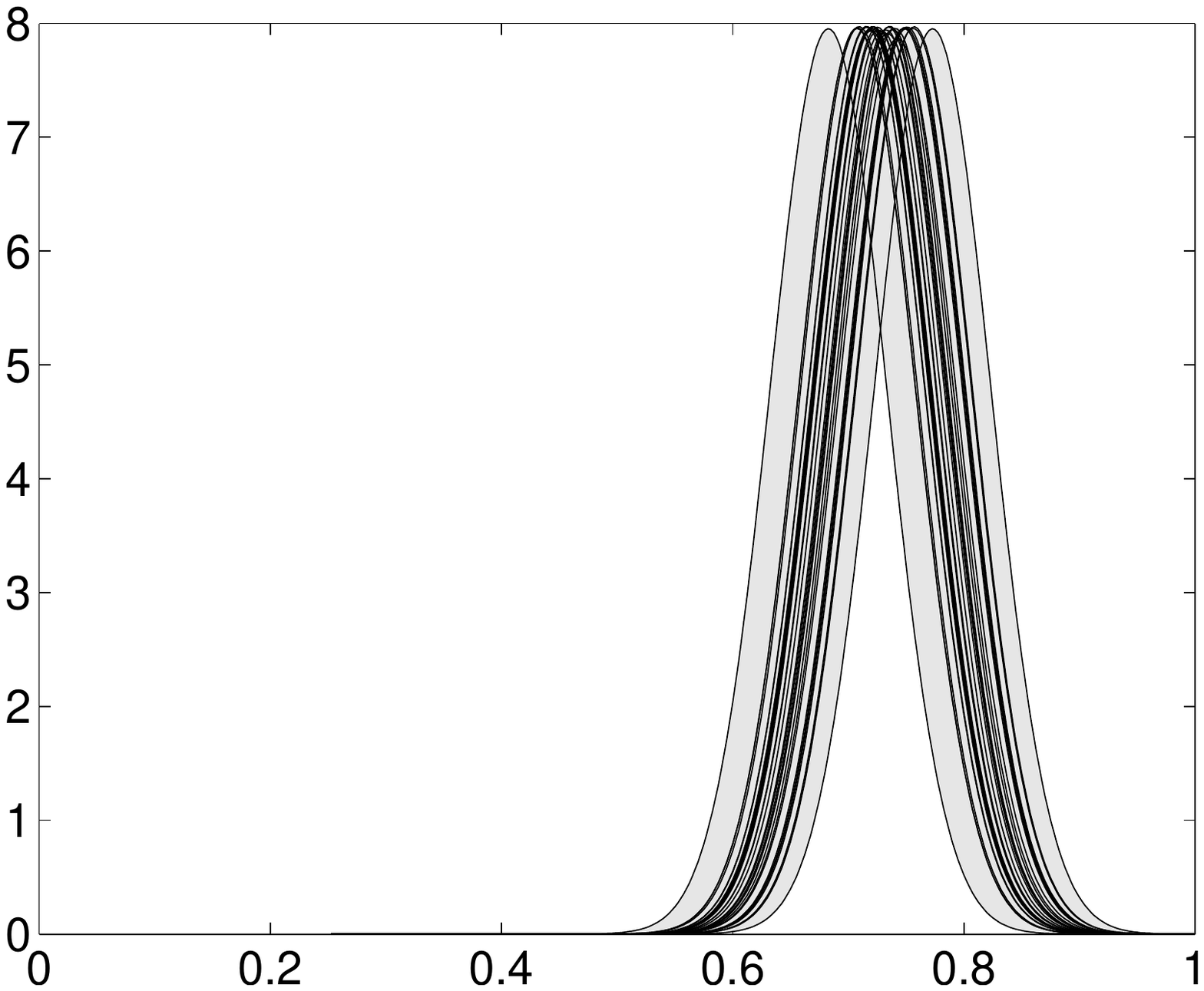}\\
$t=300$ & $t=400$\\
\includegraphics[width=0.5\textwidth, height=180pt]{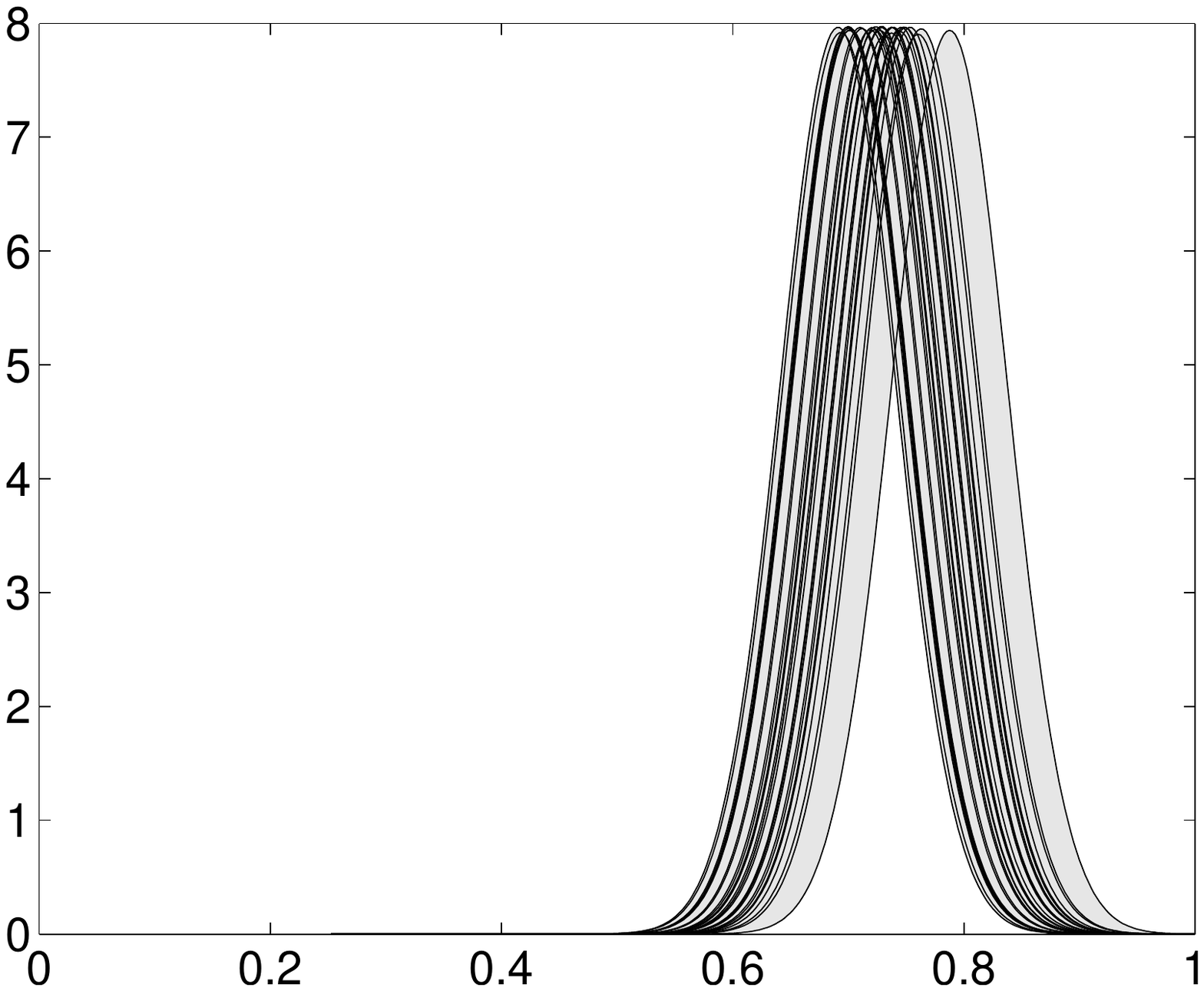}
&
\includegraphics[width=0.5\textwidth, height=180pt]{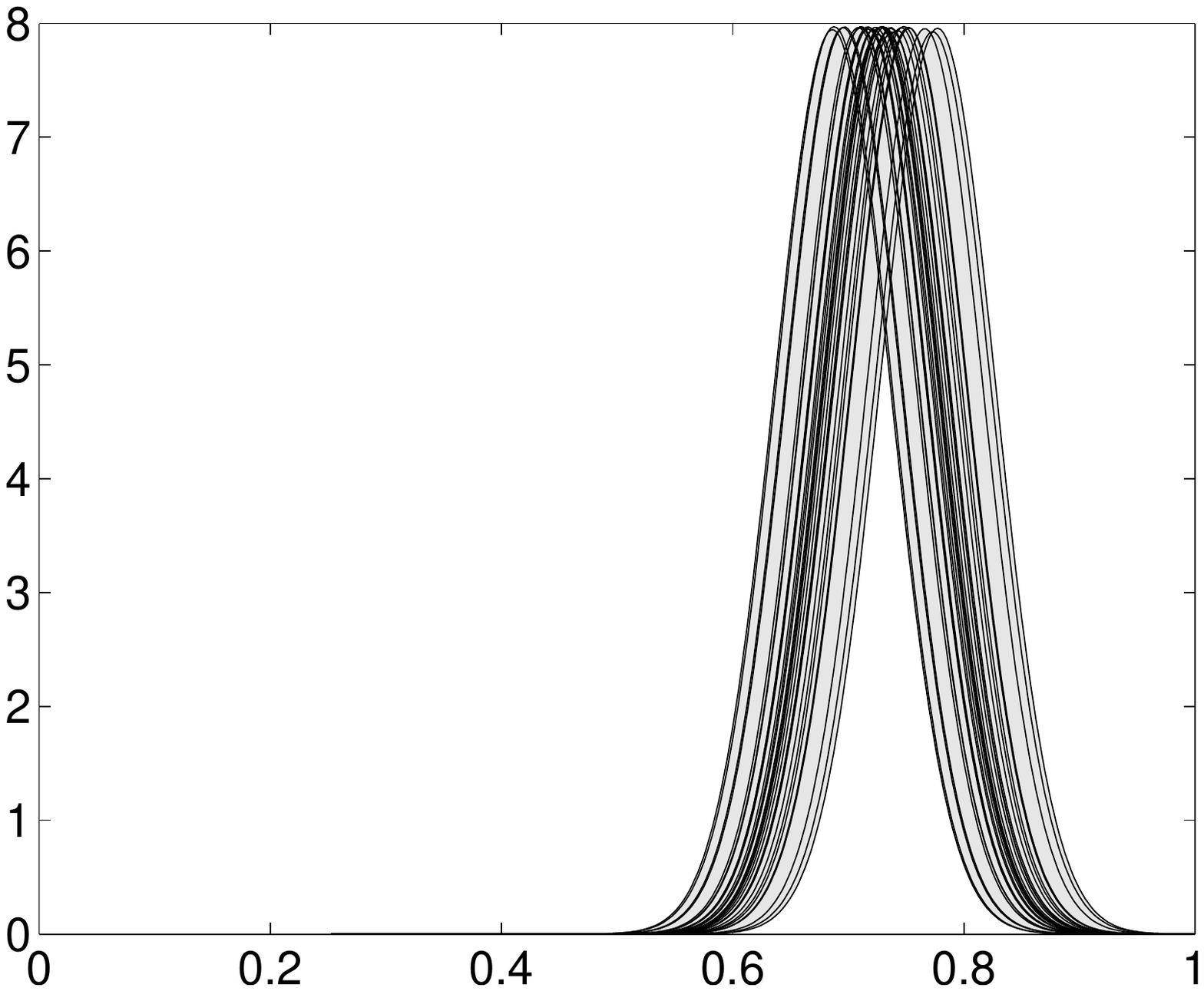}\\
\end{tabular}
\end{center}
\caption{Sequential estimation of the posterior densities of the parameter $\lambda$, for  different blocks of observations (different lines) and at different points in time $t=100,200,300,400$. For expository purposes the $\lfloor m/K \rfloor=169$ lines have been subsampled and the grey area represents the area below the envelope of the $N$ densities.}\label{Seq3}
\end{figure}

Figures \ref{Seq1}-\ref{Seq3} show an example of sequential SMCMC approximation of the posterior densities of the parameters $k$, $\nu$ and $\lambda$ for the different blocks of observations (different lines in each plot) and at different point in time $t=100,200,300,400$ (different plots) for our panel of $m=6,799$ time series. We consider $\lfloor m/M \rfloor=169$ cross-sectional blocks with $M=40$ observations each.

\end{document}